\documentclass{aa} 
\bibpunct{(}{)}{;}{a}{}{,} 

\usepackage{graphicx}
\usepackage{txfonts}
\usepackage[breaklinks,colorlinks,citecolor=black,linkcolor=black,urlcolor=black]{hyperref}

\newcommand*{\mysub}[2]{\ensuremath{#1_{\mathrm{#2}}}}
\newcommand*{\unit}[1]{\ensuremath{\mathrm{\, #1}}}

\newcommand*{\Msun}{\ensuremath{\, M_{\odot}}}
\newcommand*{\Zsun}{\ensuremath{Z_{\odot}}}
\newcommand*{\mproton}{\mysub{m}{p}}
\newcommand*{\keV}{\unit{keV}}
\newcommand*{\erg}{\unit{erg}}
\newcommand*{\cm}{\unit{cm}}
\newcommand*{\km}{\unit{km}}
\newcommand*{\Mpc}{\unit{Mpc}}
\newcommand*{\second}{\unit{s}}

\newcommand*{\Mgas}{\mysub{M}{gas}}
\newcommand*{\fgas}{\mysub{f}{gas}}
\newcommand*{\Yx}{\mysub{Y}{X}}

\newcommand*{\dA}{\mysub{d}{A}}

\newcommand*{\LCDM}{\ensuremath{\Lambda}CDM}
\newcommand*{\Omegam}{\mysub{\Omega}{m}}

\newcommand*{\Omegal}{\mysub{\Omega}{\Lambda}}
\newcommand*{\rhocr}{\mysub{\rho}{cr}}

\newcommand*{\E}[1]{\ensuremath{\times 10^{#1}}}

\newcommand*{\ltsim}{\ {\raise-.75ex\hbox{$\buildrel<\over\sim$}}\ }
\newcommand*{\gtsim}{\ {\raise-.75ex\hbox{$\buildrel>\over\sim$}}\ }
\newcommand*{\proptosim}{\ {\raise-.75ex\hbox{$\buildrel\propto\over\sim$}}\ }

\newcommand*{\Chandra}{{\it Chandra}}
\newcommand*{\Planck}{{\it Planck}}

\defcitealias{Pacaud1512.04264}{II}
\newcommand*{\xxlii}{\citetalias{Pacaud1512.04264}}
\defcitealias{Giles1512.03833}{III}
\newcommand*{\xxliii}{\citetalias{Giles1512.03833}}
\defcitealias{Lieu1512.03857}{IV}
\newcommand*{\xxliv}{\citetalias{Lieu1512.03857}}
\defcitealias{Mantz1401.2087}{V}
\newcommand*{\xxlv}{\citetalias{Mantz1401.2087}}
\defcitealias{Eckert1512.03814}{XIII}
\newcommand*{\xxlxiii}{\citetalias{Eckert1512.03814}}

\newcommand*{\cl}{XLSSC\,122}
\newcommand*{\clfull}{3XLSS\,J021744.0$-$034531}

\newcommand*{\figscale}{0.9}

\begin{document}

\title{The XXL Survey: XVII. X-ray and Sunyaev-Zel'dovich Properties of the Redshift 2.0 Galaxy Cluster \cl{}\thanks{These results are based on observations obtained with XMM-{\it Newton}, an ESA science mission with instruments and contributions directly funded by ESA Member States and NASA; the \Chandra{} X-ray Observatory; and the Combined Array for Research in Millimeter-wave Astronomy (CARMA).}}
\titlerunning{The XXL Survey. XVII}

\author{A.~B.~Mantz,\inst{\ref{kipac}}\fnmsep\inst{\ref{stanford}}\thanks{E-mail: \href{mailto:amantz@slac.stanford.edu}{\tt amantz@slac.stanford.edu}}
  Z.~Abdulla,\inst{\ref{kicp}}\fnmsep\inst{\ref{uchicagoaa}}
  S.~W.~Allen,\inst{\ref{kipac}}\fnmsep\inst{\ref{stanford}}\fnmsep\inst{\ref{slac}}
  J.~E.~Carlstrom,\inst{\ref{kicp}}\fnmsep\inst{\ref{uchicagoaa}}\fnmsep\inst{\ref{uchicagoph}}
  C.~H.~A.~Logan,\inst{\ref{bristol}}
  D.~P.~Marrone,\inst{\ref{steward}}
  B.~J.~Maughan,\inst{\ref{bristol}}
  J.~Willis,\inst{\ref{uvic}}
  F.~Pacaud,\inst{\ref{bonn}}
  M.~Pierre\inst{\ref{cea}}
}
\institute{
  Kavli Institute for Particle Astrophysics and Cosmology, Stanford University, 452 Lomita Mall, Stanford, CA 94305, USA \label{kipac}
  \and
  Department of Physics, Stanford University, 382 Via Pueblo Mall, Stanford, CA 94305, USA \label{stanford}
  \and
  Kavli Institute for Cosmological Physics, University of Chicago, 5640 South Ellis Avenue, Chicago, IL 60637, USA\label{kicp}
  \and
  Department of Astronomy and Astrophysics, University of Chicago, 5640 South Ellis Avenue, Chicago, IL 60637, USA\label{uchicagoaa}
  \and
  SLAC National Accelerator Laboratory, 2575 Sand Hill Road, Menlo Park, CA  94025, USA\label{slac}
  \and
  Department of Physics/Enrico Fermi Institute, University of Chicago, 5640 South Ellis Avenue, Chicago, IL 60637, USA\label{uchicagoph}
  \and
  H. H. Wills Physics Laboratory, University of Bristol, Tyndall Avenue, Bristol BS8 1TL, UK\label{bristol}
  \and
  Steward Observatory, University of Arizona, 933 North Cherry Avenue, Tucson, AZ 85721, USA\label{steward}
  \and
  Department of Physics and Astronomy, University of Victoria, 3800 Finnerty Road, Victoria, BC, Canada\label{uvic}
  \and
  Argelander-Institute for Astronomy, Auf dem H\"ugel 71, D-53121 Bonn, Germany\label{bonn}
  \and
  Service d'Astrophysique AIM, DRF/IRFU/SAp, CEA-Saclay, 91191 Gif-sur-Yvette, France\label{cea}
}
\authorrunning{A. B. Mantz et al.}

\date{Submitted Nov 18, 2016. Accepted March 21, 2017.}

\abstract{
  We present results from a 100\,ks XMM-{\it Newton} observation of galaxy cluster \cl{}, the first massive cluster discovered through its X-ray emission at $z\approx2$. The data provide the first precise constraints on the bulk thermodynamic properties of such a distant cluster, as well as an X-ray spectroscopic confirmation of its redshift. We measure an average temperature of $kT=5.0\pm0.7$\,keV; a metallicity with respect to solar of $Z/\Zsun=0.33^{+0.19}_{-0.17}$, consistent with lower-redshift clusters; and a redshift of $z=1.99^{+0.07}_{-0.06}$, consistent with the earlier photo-$z$ estimate. The measured gas density profile leads to a mass estimate at $r_{500}$ of $M_{500}=(6.3\pm1.5)\E{13}\Msun$. From CARMA 30\,GHz data, we measure the spherically integrated Compton parameter within $r_{500}$ to be $Y_{500}=(3.6\pm0.4)\E{-12}$. We compare the measured properties of \cl{} to lower-redshift cluster samples, and find good agreement when assuming the simplest (self-similar) form for the evolution of cluster scaling relations. While a single cluster provides limited information, this result suggests that the evolution of the intracluster medium in the most massive, well developed clusters is remarkably simple, even out to the highest redshifts where they have been found. At the same time, our data reaffirm the previously reported spatial offset between the centers of the X-ray and SZ signals for \cl{}, suggesting a disturbed configuration. Higher spatial resolution data could thus provide greater insights into the internal dynamics of this system.
}


\maketitle

\section{Introduction} \label{sec:intro}

Galaxy clusters at redshifts $z>1$ are now routinely being discovered in surveys at X-ray (\citealt{Pacaud0709.1950}; \citealt{Pacaud1512.04264}, hereafter XXL Paper~\xxlii{}; \citealt{Willis1212.4185}), IR \citep{Papovich1002.3158, Gobat1011.1837, Brodwin1205.3787, Brodwin1410.2355, Brodwin1504.01397, Stanford1205.3786, Zeimann1207.4793}, and mm \citep{Hasselfield1301.0816, Bleem1409.0850} wavelengths. This includes a small but growing number of clusters (and protoclusters) at $z>1.75$ \citep{Andreon0812.1699, Andreon1311.4361, Gobat1011.1837, Spitler1112.2691, Brodwin1504.01397, Hung1605.07176}, corresponding to lookback times $\gtsim10$\,Gyr, raising the exciting possibility that statistical studies of the cluster population during their epoch of formation may not be far away. At present, however, the great majority of these $z\gg1$ clusters have been identified as overdensities of IR-luminous galaxies. Compared to samples selected on signals from the intracluster medium (ICM), these clusters are less likely to represent well developed, approximately virialized halos, complicating comparisons to the best-studied cluster samples at lower redshifts. This feature also increases the challenge of precisely characterizing clusters at these high redshifts using observations of the ICM, since IR-selected clusters present fainter X-ray emission and a weaker Sunyaev-Zel'dovich (SZ) effect (e.g.\ \citealt{Culverhouse1007.2853}). Indeed, excluding the present work, characterization of the ICM properties of known clusters at $z\gtsim2$ has been limited to simple X-ray flux measurements.

This paper concerns \cl{} (formally \clfull{}), the highest-redshift confirmed galaxy cluster detected in the XMM-{\it Newton} Large Scale Structure survey (XMM-LSS), as well as in its extension, the XMM-XXL survey \citep{Pierre0305191, Pierre1512.04317}. On the basis of follow-up imaging spanning the optical and IR spectrum, \citet{Willis1212.4185} assigned \cl{} a photometric redshift of $z=1.9^{+0.19}_{-0.21}$. Subsequent 30\,GHz continuum observations with CARMA provided a significant detection of the SZ effect towards the cluster, confirming the presence of a hot ICM (\citealt{Mantz1401.2087}, hereafter XXL Paper~\xxlv{}). While the X-ray flux and Compton $Y$ parameter of \cl{} were shown to be broadly consistent with lower-redshift clusters under simple evolutionary assumptions, the XMM survey detection \citep{Willis1212.4185, Clerc1408.6325} provided only $\sim100$ source counts, too few to measure more detailed X-ray properties. Here we present results from a deeper XMM observation of \cl{}, which allows us to, for the first time, obtain precise constraints on the gas mass, average temperature and metallicity of a massive cluster at $z\approx2$, as well as an X-ray spectroscopic confirmation of its redshift.\footnote{To date, we have been unable to verify the cluster redshift through optical/NIR spectroscopy of cluster members. However, NIR spectroscopic data for bright sources in this field that are not associated with the cluster are presented by Adami et~al.\ (XXL Paper XX, in preparation).} We also present the analysis of a short \Chandra{} observation of \cl{}, as part of a program to quantify the level of active galactic nucleus (AGN) contamination for the X-ray signal from $z>1$ XMM-LSS  cluster candidates. In addition, we update the SZ effect measurements of \cl{}, incorporating CARMA data that were obtained after the initial reported detection. A companion paper by Horellou et al.\ (in preparation) presents Herschel IR and LABOCA sub-mm data covering \cl{} and discusses star formation in this exceptional cluster.

Throughout this work, we assume a concordance \LCDM{} cosmological model, with dark energy in the form of a cosmological constant, described by Hubble parameter $H_0=70\km\second^{-1}\Mpc^{-1}$, matter density $\Omegam=0.3$ and dark energy density $\Omegal=0.7$; in this model, at $z=2$, a projected distance of $1''$ corresponds to 8.37\,kpc. Quoted uncertainties refer to 68.3\% confidence intervals. We report dimensionless, spherically integrated Compton parameter ($Y$) in units of steradians.

Section~\ref{sec:data} details the X-ray and SZ data used here, and their reduction. In Section~\ref{sec:results}, we present the results of our X-ray imaging and spectroscopic analysis, as well as the SZ data analysis, and the measurements of the redshift and global thermodynamic properties of \cl{}. In Section~\ref{sec:discussion}, we compare the scaling properties of \cl{} with well studied cluster samples at lower redshifts, discuss implications of its metallicity, and comment on its dynamical state. We conclude in Section~\ref{sec:conclusion}.

\section{Data Reduction} \label{sec:data}

\subsection{XMM-Newton} \label{sec:xmmdata}

The deep XMM observation, ObsID 0760540101, was obtained on 16 July, 2015. Our X-ray analysis is based on only this new observation, since \cl{} falls far off-axis in all of the previous exposures. The data were reduced using the XMM-{\it Newton} Extended Source Analysis Software ({\sc xmm-esas}; version 15.0.0),\footnote{\url{http://www.cosmos.esa.int/web/xmm-newton/sas}} following the recommendations of \citet{Snowden0710.2241} and the {\sc xmm-esas} Cookbook.\footnote{\url{http://heasarc.gsfc.nasa.gov/docs/xmm/esas/cookbook/}} Following standard calibration and filtering of the raw event files, lightcurves for each of the EPIC detectors were manually inspected, and a period of approximately 18\,ks at the end of the observation during which the X-ray background was enhanced was manually removed. There is no indication that any of the functioning MOS CCDs were in an anomalous state during the observation. The quiescent particle background (QPB) model generated by XMM-ESAS is spectrally consistent with the observed data at all energies $\gtsim6$\,keV, indicating that the particle background was well behaved during the cleaned exposure. The final clean exposure times for PN, MOS1 and MOS2 are, respectively, 73.4, 84.1 and 83.5\,ks.

Although \cl{} is resolved in the XMM data, the point spread function (PSF) still has an impact on the observed cluster emission. In our imaging analysis and deprojection (Sections~\ref{sec:image} and \ref{sec:deproj}), we account for this effect using a symmetric version of the Gaussian+beta model of the EPIC PSF from \citet{Read1108.4835}. In the determination of the deprojected density profile, this correction exceeds the $1\sigma$ statistical uncertainties only in the cluster center ($r<100$\,kpc). When fitting models to either the image or spectral data, we use the appropriate Poisson likelihood for the observed counts.

\subsection{Chandra} \label{sec:chandradata}

An 11\,ks observation of \cl{} was obtained with \Chandra{} (ObsID 18263) in order to constrain the level of any AGN emission that could have contributed to the original XMM detection. The data were reprocessed in the standard way using {\sc ciao}\footnote{\url{http://cxc.harvard.edu/ciao/}} version 4.8 and {\sc caldb}\footnote{\url{http://cxc.harvard.edu/caldb/}} version 4.7.2. No periods of background flaring were present in the data. In this work, we perform a preliminary analysis of these data; a complete analysis of the AGN contamination for full sample of high-$z$ XMM-LSS clusters will be presented in future work.

\subsection{CARMA} \label{sec:carmadata}

XXL Paper~\xxlv{} presents 30\,GHz continuum observations of the SZ effect towards \cl{}, measured with the 8-element array of 3.5\,m CARMA antennas (hereafter CARMA-8). The 3.5\,m antennas were arranged with 6 elements in a compact configuration and two outlying elements, respectively providing baselines corresponding to $uv$ radii of 0.3--9.7\,k$\lambda$. The signals were processed by the CARMA wideband (WB) correlator in sixteen 500\,MHz sub-bands, each consisting of 16 channels. These data were obtained between March 2012 and November 2013, with most of the observations being in those two months. The effective on-source exposure time (after data flagging) for the CARMA-8 data is 52.8\,h.

Here we add to this data set observations that used the full 23-element CARMA array, comprising six 10.4\,m, nine 6.1\,m, and eight 3.5\,m antennas (hereafter CARMA-23).\footnote{In fact, two of the 10.4\,m elements and the two outlying 3.5\,m elements were offline during these observations, significantly reducing our sensitivity at long baselines, but we nevertheless adhere to the convention of calling these CARMA-23 data.} These data were obtained in July 2014. The 10.4\,m and 6.1\,m antennas were arranged in the CARMA E configuration, with the 6.1\,m elements maximally compact and the 10.4\,m elements located around the periphery. The 3.5\,m antennas were arranged as described above. The WB correlator was used to process signals from 8 of the 6.1\,m antennas, since these provide the maximum collecting area at the short baselines where the cluster SZ signal is strongest, while the CARMA spectral line (SL) correlator processed 2\,GHz of bandwidth from all 23 elements (with baselines also represented in the WB data removed in later analysis). The central frequency of the observations was 30.938\,GHz, for which the array samples $uv$ radii of 0.3--7.5\,k$\lambda$. The large range of baselines probed in both the CARMA-8 and CARMA-23 data critically allows the flux of contaminating, point-like radio galaxies to be measured contemporaneously with the cluster signal, with the latter being strongest on spatial scales corresponding to $uv$ radii $\ltsim2$\,k$\lambda$. Within the CARMA-23 data set, the WB data are most sensitive to galaxy cluster scales (short baselines), while the SL data are most sensitive to compact sources (long baselines). The effective on-source exposure times for the WB and SL data are, respectively, 4.3\,h and 1.0\,h.

The data reduction procedure for CARMA-8 observations is described by \citet{Muchovej0610115}; this includes flagging for weather, shadowing and technical issues, as well as bandpass and gain calibration using observations of bright quasars interleaved with the cluster observations. The reduction of the CARMA-23 data from a given correlator follows the same procedure. The absolute flux calibration is tied to the Mars model of \citet{Rudy1987PhDT.........4R}, which is accurate to better than 5 per cent. A common flux calibration for all antennas is obtained using periodic observations of Mars with the compact 3.5\,m antennas (for which Mars is always unresolved), and bootstrapping to all antennas using observations of unresolved, bright quasars. Table~\ref{tab:carma} summarizes the noise levels and synthesized beams of the CARMA observations.

\begin{table}
  \begin{center}
    \caption{
      CARMA RMS map noise and synthesized beam shapes.
    }
    \label{tab:carma}
    \vspace{-1ex}
    \begin{tabular}{lcccc}
      \hline\hline
      Array & Noise & $a$ & $b$ & $\phi$ \\
      & (mJy/beam) & ($''$) & ($''$) & (deg) \\
      \hline
      \multicolumn{5}{c}{$uv$ radii $<2$\,k$\lambda$} \\
      CARMA-8 & 0.15 & 117 & 132 & $-1$ \\
      CARMA-23 (WB) & 0.30 & 69 & 92 & $-19$ \\
      CARMA-23 (SL) & 0.53 & 65 & 66 & $-50$ \\
      \multicolumn{5}{c}{$uv$ radii $>2$\,k$\lambda$} \\
      CARMA-8 & 0.17 & 18 & 25 & $+35$ \\
      CARMA-23 (WB) & 0.50 & 29 & 89 & $-13$ \\
      CARMA-23 (SL) & 0.21 & 24 & 27 & $-25$ \\
      \hline
    \end{tabular}
    \tablefoot{
      Beam shapes are given as full-width-half-maximum minor and major axes, and position angles east from north of the major axes. Here the visibility data are divided into $uv$ radii $<2$\,k$\lambda$ and $>2$\,k$\lambda$; the former provide most of the sensitivity to the cluster SZ signal, while the latter are predominantly useful for constraining point-like radio emission in the field. (In practice, all baselines are calibrated and analyzed simultaneously.)
    }
  \end{center}
\end{table}

\section{Results} \label{sec:results}

\subsection{XMM Image} \label{sec:image}

\begin{figure}
  \centering
  \includegraphics[scale=0.32]{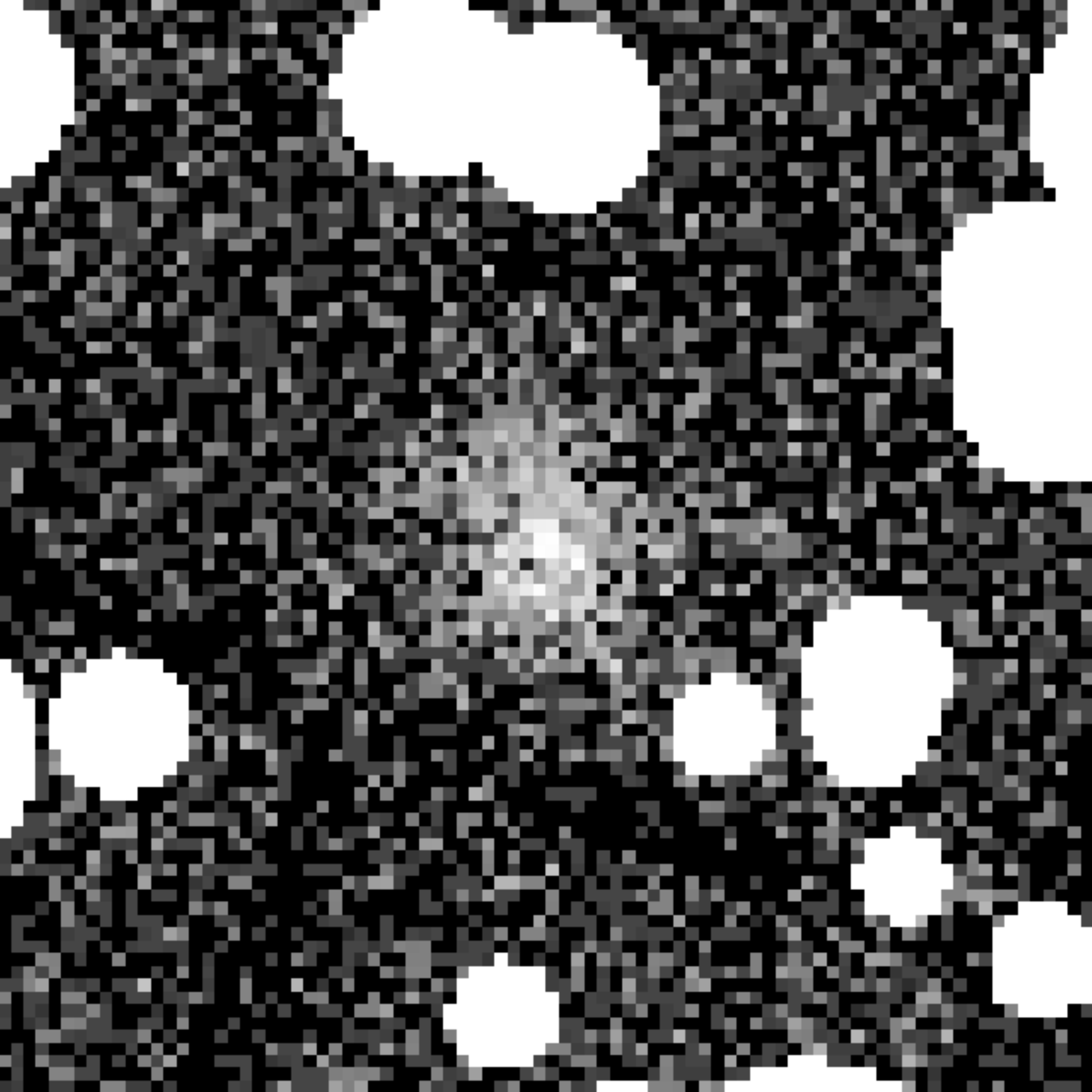}
  \caption{
    0.4--3.0\,keV image of \cl{}, combining the MOS and PN data, with nearby point sources masked. Note that some artifacts due to the PN chip gaps are visible. The scale is the same as in the left panel of Figure~\ref{fig:image}.
  }
  \label{fig:imageu}
\end{figure}

\begin{figure*}
  \centering
  \includegraphics[scale=0.32]{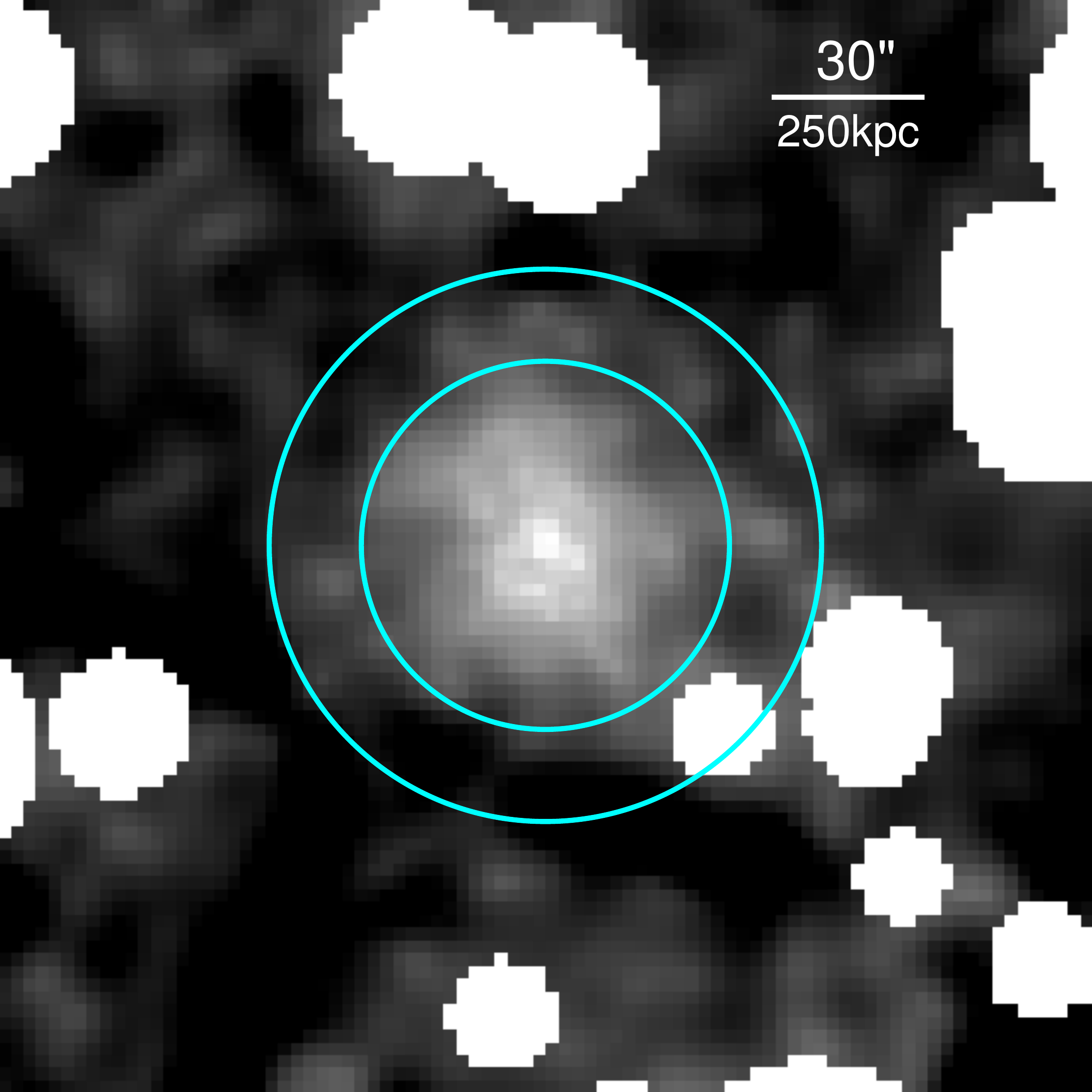}
  \hspace{10mm}
  \includegraphics[scale=0.32]{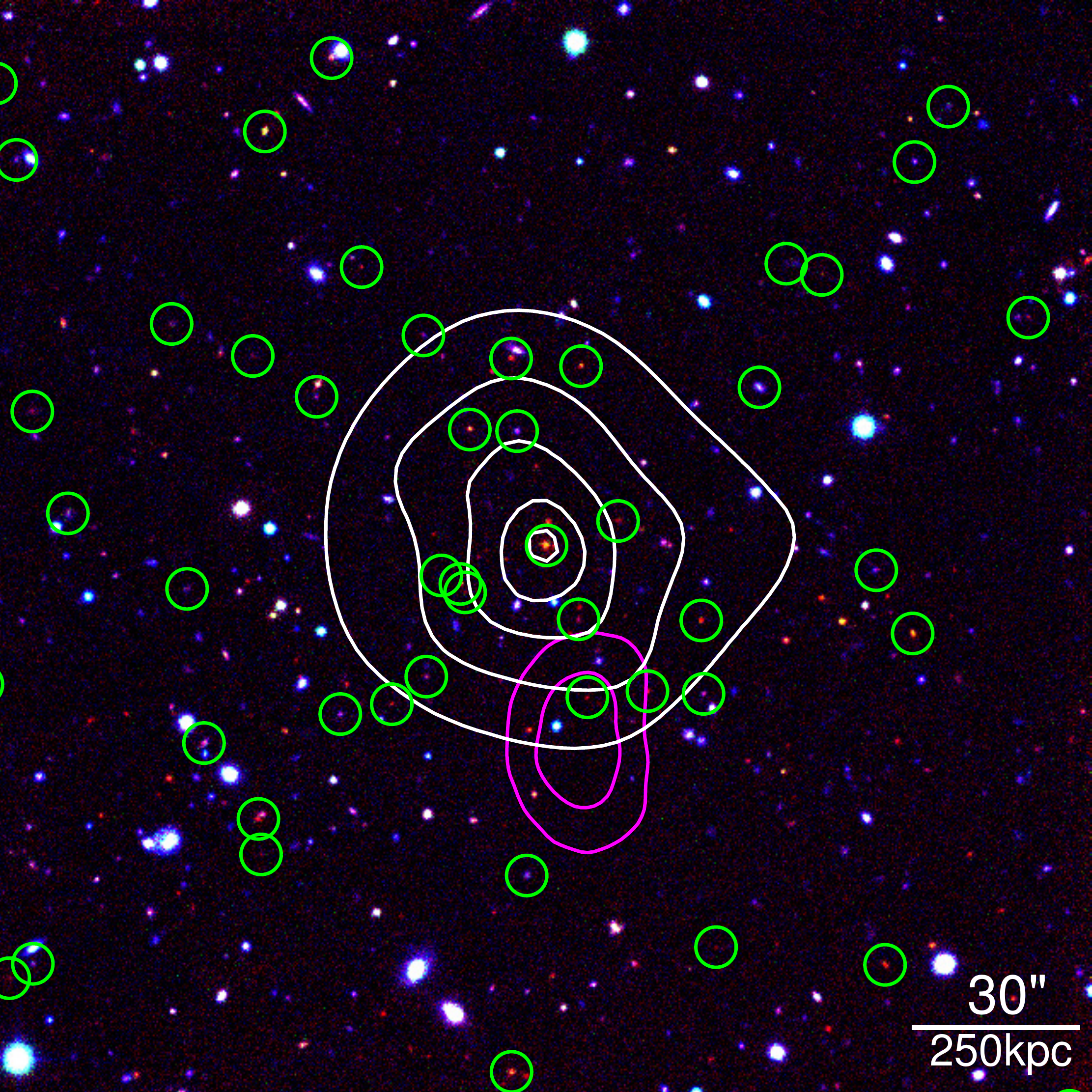}
  \caption{
    Left: adaptively smoothed 0.4--3.0\,keV image of \cl{}, combining the MOS and PN data, with nearby point sources masked. Note that some artifacts due to the PN chip gaps remain visible. Cyan circles show our estimates of $r_{500}$ and $r_{200}$ from Section~\ref{sec:global}.
    Right: $iJK$ image with smoothed X-ray surface brightness contours overlaid in white. The outermost contour approximately corresponds to $r_{500}$. Galaxies that were photometrically identified with the cluster redshift (photo-$z$'s between 1.7 and 2.1) by \citet{Willis1212.4185} are circled in green. Magenta contours show the 68.3 and 95.4 per cent confidence regions for the center of a symmetric cluster model fit to the CARMA data (see Section~\ref{sec:sz}).
  }
  \label{fig:image}
\end{figure*}

A combined EPIC (PN+MOS1+MOS2), 0.4--3.0\,keV image of \cl{} is shown in Figure~\ref{fig:imageu}, and an adaptively smoothed version appears in the left panel of Figure~\ref{fig:image}.\footnote{Given the high expected redshift of the cluster, we expect negligible emission at observer-frame energies $\gtsim3$\,keV, as confirmed in Section~\ref{sec:spectra}.} The right panel of Figure ~\ref{fig:image} compares contours of X-ray surface brightness with the corresponding $iJK$ image, with potential cluster members (galaxies with photo-$z$'s of 1.7--2.1) circled \citep{Willis1212.4185}. The positions of the X-ray brightness peak, the X-ray centroid, and the putative brightest cluster galaxy (BCG; the brightest likely cluster member from \citealt{Willis1212.4185} in $K$ band) all coincide to within $\sim1''$ (the MOS pixel size). We adopt the BCG position, J02:17:44.190$-$03:45:31.46, as the cluster center in subsequent analysis. Contaminating X-ray point sources are masked in the displayed image and removed from our analysis in general. Note that none of these X-ray point sources is coincident with a photometrically identified cluster member.

\begin{figure}
  \centering
  \includegraphics[scale=\figscale]{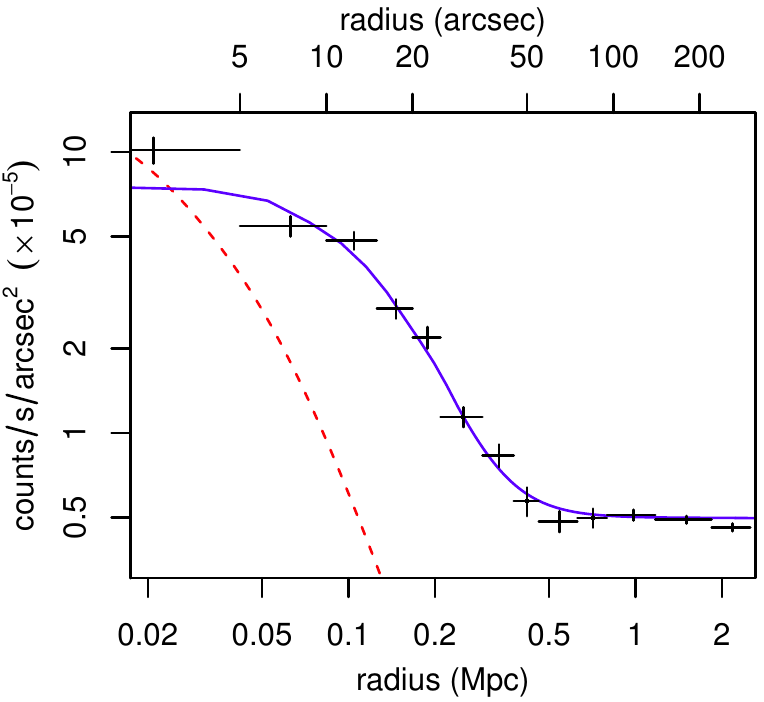}
  \caption{
    QPB-subtracted surface brightness profile measured in the 0.4--3.0\,keV band. The best-fitting beta model plus constant, after convolution with the PSF, is shown as a solid, blue line. The profile of the PSF, arbitrarily normalized, is shown as a dashed, red line.
  }
  \label{fig:sbprof}
\end{figure}

Figure~\ref{fig:sbprof} shows a binned, QPB-subtracted surface brightness profile extracted from the X-ray image. We model the observed counts as the sum of the QPB and astrophysical components, where the astrophysical model is the sum of a constant background and a beta-model cluster surface brightness, transformed appropriately by the point spread function and the exposure map. The QPB-subtracted surface brightness corresponding to the best-fitting model is shown in the figure. Comparing the observed surface brightness with the constant component of this fit, we estimate that the cluster signal exceeds the background at radii $\ltsim35''$, with $\sim950$ source counts in the 0.4--3.0\,keV band falling within this radius.

\subsection{Chandra Limits on Point-Source Contamination} \label{sec:chandrares}

The modest spatial resolution of XMM makes the identification and removal of point sources such as AGN difficult for distant clusters. In contrast, \Chandra's higher resolution makes it straightforward to identify point-like emission, even given its smaller effective area. For this reason, we obtained an 11\,ks \Chandra{} observation of \cl{} -- too short to detect the diffuse emission from the ICM, but sufficient to place interesting constraints on emission from discrete point sources in the field. No point sources are detected in the \Chandra{} data within $35''$ of the BCG position either using the {\sc ciao}
detection tools ( {\sc celldetect}, {\sc vtpdetect} and {\sc wavdetect}), or by eye.

We estimated an upper limit on the 0.5--2.0\,keV flux of a hypothetical point source at the position of the BCG, using the {\sc ciao} tool {\sc aprates}\footnote{http://cxc.harvard.edu/ciao/threads/aprates/}. In order to convert photon counts in the 0.5-2.0\,keV band to flux, we assumed an absorbed power-law spectrum and an equivalent absorbing hydrogen column density of $2.02\times 10^{20}\cm^2$
(Kalberla et al 2005). The resulting 68.3 and 95.4 per cent confidence upper limits respectively correspond to 8 and 20 per cent of the 0.5--2.0\,keV flux measured within $r_{500}$ from the XMM data (see Section~\ref{sec:global}), and are insensitive to the precise choice of photon index (in the range of 1.4--1.9) at the per cent level. This constraint is somewhat tighter than (and consistent with) the limits on contamination from unresolved AGN that we obtain spectrally in the next section.

\subsection{XMM Spectral Analysis} \label{sec:spectra}

\begin{figure*}
  \centering
  \includegraphics[scale=0.96]{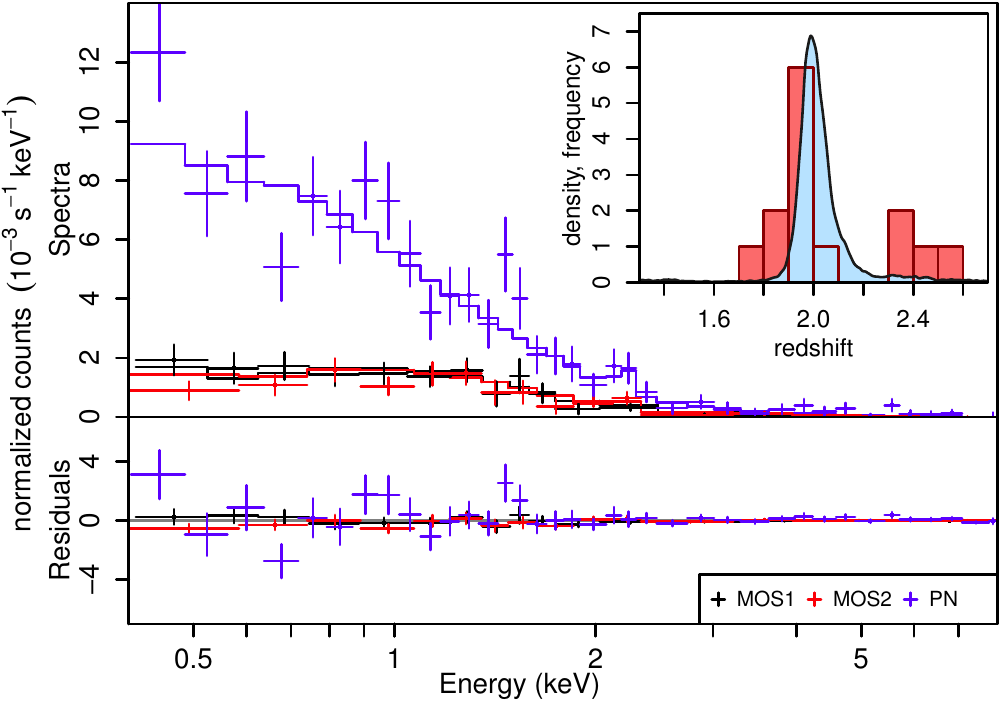}
  \hspace{3mm}
  \includegraphics[scale=0.96]{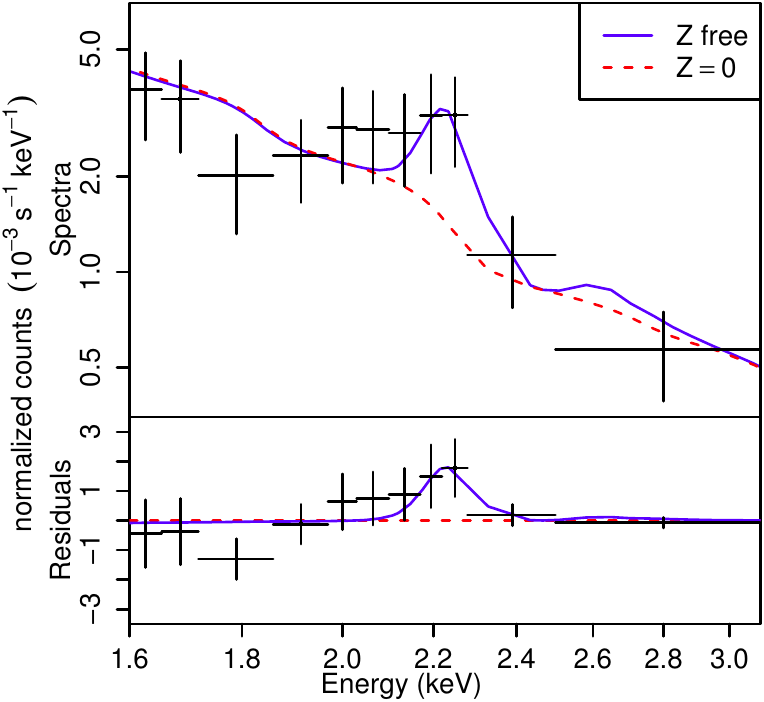}
  \caption{
    Left: background-subtracted spectra from each EPIC detector extracted from a circle of radius $35''$ centered on the BCG. Solid lines show the best fitting folded model, consisting of a redshifted thermal emission spectrum with Galactic absorption (temperature, metallicity, redshift and normalization are all free here). The inset shows the posterior distribution of the cluster redshift from this fit (blue shading), compared with the photometric redshift histogram for galaxies within $30''$ of the X-ray center from \citet{Willis1212.4185}. (The full photo-$z$ histogram includes projections at lower and higher redshifts; we show only the vicinity of the X-ray constraint here.) The X-ray redshift constraint is possible due to the detection of the rest-frame 6.7\,keV Fe emission line complex at a redshifted energy of $\sim2.2$\,keV.
    Right: the combined EPIC spectrum is compared with the best-fitting model (solid, blue curve) in the vicinity of the Fe emission complex. The dashed, red curve shows the same model with zero metallicity. Residuals are relative to the $Z=0$ model. The Fe emission feature is formally detected at $2.6\sigma$ significance.
  }
  \label{fig:spectra}
\end{figure*}

To investigate the spectral properties of \cl{}, we focus on the region identified in Section~\ref{sec:image} where the cluster surface brightness exceeds the astrophysical background, a circle of radius $35''$ (293\,kpc) centered on the BCG. We note that this radius is conveniently very close to the estimate of $r_{500}$ arrived at in Section~\ref{sec:global}. For each EPIC instrument, we generated spectra and response matrices for this region, as well as for an annulus spanning radii of $2.3'$--$5'$ (1.16--2.51\,Mpc; the outermost two bins in Figure~\ref{fig:sbprof}) to serve as an estimate of the local background. All spectra were grouped to have at minimum one count per channel. The background-subtracted spectra of the cluster are shown in the left panel of Figure~\ref{fig:spectra}.

Our spectral analysis is performed using {\sc xspec}\footnote{\url{http://heasarc.gsfc.nasa.gov/docs/xanadu/xspec/}} (version 12.9.0o). We model thermal emission from the ICM as a sum of bremsstrahlung continuum and line emission components, evaluated using the {\sc apec} plasma model (ATOMDB version 2.0.2). Relative metal abundances were fixed to the solar ratios of \citet{Anders1989GeCoA..53..197A}, with the overall metallicity allowed to vary. Photoelectric absorption by Galactic gas was accounted for using the {\sc phabs} model, employing the cross sections of \citet{Balucinska1992ApJ...400..699B}, and adopting a fixed value of $2.02\E{20}\cm^{-2}$ for the equivalent absorbing hydrogen column density \citep{Kalberla0504140}. We fit models using the \citet{Cash1979ApJ...228..939} statistic, as modified by \citet[][the $C$ statistic]{Arnaud1996ASPC..101...17A}, to properly account for the Poisson nature of the source and background counts. Confidence regions were determined by Markov Chain Monte Carlo  explorations of the model parameter spaces.\footnote{We use the {\sc lmc} code: \url{https://github.com/abmantz/lmc}}

We perform a series of tests using the full energy band shown in the left panel of Figure~\ref{fig:spectra}, 0.4--8.0\,keV. First, we test for consistency of the PN and MOS responses by fitting a thermal emission model with Galactic absorption, with the cluster temperature, metallicity and redshift free, allowing the different normalizations to apply to the PN and MOS1+MOS2 detectors. These two normalizations are consistent at the $1\sigma$ level, and we henceforth assume a single, linked normalization for all detectors. The resulting fit, shown in the figure, has $C=879.4$ for 1020 degrees of freedom; this corresponds to the 43rd percentile of $C$ values obtained from random data sets generated from the best-fitting model, indicating an acceptable goodness of fit.

We next test for the presence of AGN emission in the extracted cluster spectra, beyond that accounted for by the local background measurement. Multiple lines of evidence indicate that significant point source contamination is unlikely to be present, namely the lack of point-like emission detected within the cluster in the CARMA and \Chandra{} data (Sections~\ref{sec:carmadata} and \ref{sec:chandrares}), and the consistency of the BCG's $J-K$ color with a passively evolving galaxy at $z\sim1.9$ \citep{Willis1212.4185}. Nevertheless, we test for the possibility of residual AGN contamination spectrally, by introducing a power-law emission component to the model. Given the lack of discrete sources detected by \Chandra{}, we adopt a fixed photon index of 1.4, appropriate for a background population of unresolved AGN, and compare the best-fitting $C$ statistic value for this model with that of the cluster-only model using an $F$ test. The resulting significance value is 0.16, indicating that there is statistically no improvement to the fit from including a power-law component. Conversely, constraints on the normalization of the power-law model can be translated into limits on the fraction of the total flux that could be due to AGN contamination. At observer-frame energies of 0.5--2.0\,keV, the primary band used for cluster detection in the XXL survey \citep{Pacaud0607177}, the $1\sigma$ and $2\sigma$ upper limits on possible AGN contamination are, respectively, 19 and 30 per cent. Since it is statistically disfavored, and because of the considerations mentioned above, we do not include a power-law component in subsequent analysis, but we note that the constraints on the cluster temperature, metallicity and redshift arrived at in both cases are compatible (Table~\ref{tab:specfits}).

\begin{table*}
  \begin{center}
    \caption{
      Results of X-ray spectral fits to the \cl{} data within a radius of $35''$.
    }
    \label{tab:specfits}
    \vspace{-1ex}
    \begin{tabular}{ccccccc}
      \hline\hline\vspace{-2ex}\\
      Band (keV) & AGN & $C$ & dof & $kT$ (keV) & $Z$ (solar) & $z$\vspace{0.3ex}\\
      \hline\vspace{-2ex}\\
      0.4--8.0 & N & 879.4 & 1020 & $5.1^{+0.9}_{-0.5}$ & $0.32\pm0.17$ & $1.99^{+0.07}_{-0.05}$\vspace{0.5ex}\\
      0.4--8.0 & Y & 877.8 & 1019 & $4.1^{+1.2}_{-0.8}$ & $0.36_{-0.22}^{+0.25}$ & $1.99^{+0.07}_{-0.05}$\vspace{0.5ex}\\
      0.4--8.0 & N & 887.5 & 1022 & $5.8^{+0.7}_{-0.6}$ & 0.00 & 1.99\medskip\\
      {\bf 0.4--3.0} & {\bf N} & {\bf 606.5} & {\bf 721} & $\mathbf{5.0\pm0.7}$ & $\mathbf{0.33^{+0.19}_{-0.17}}$ & $\mathbf{1.99^{+0.07}_{-0.05}}$\vspace{0.5ex}\\
      \hline
   \end{tabular}
   \tablefoot{
     [1] Energy range used in the fit; [2] whether a power-law emission with a photon index of 1.4 component was included in the model; [3--4] modified Cash statistic corresponding to the best fit, and number of degrees of freedom for the model; [5--7] best-fitting values and 68.3 per cent confidence intervals for the cluster temperature, metallicity and redshift. Quantities without error bars were held fixed.
   }
  \end{center}
\end{table*}

An emission feature is clearly visible in the PN spectrum (Figure~\ref{fig:spectra}) at $\sim2.2$\,keV. Given the presence of hot gas in the cluster (XXL Paper~\xxlv{}) and the photometric redshift of $z=1.9^{+0.19}_{-0.21}$ from \citet{Willis1212.4185}, we identify this feature as the rest-frame 6.7\,keV Fe emission line complex, redshifted by a factor of $\sim3$.\footnote{Note that it is unlikely that this feature is due to instrumental or astrophysical backgrounds. While there is an Au fluorescence line at 2.2\,keV, it is typically faint and observed most easily in MOS data, whereas the observed line is clearly visible in the PN spectrum. Likewise, solar wind charge exchange typically does not produce lines at such a high energy, and we would expect to see even more prominent emission lines at lower energies in that case. Furthermore, even if they were present, both of these cases are in principle handled by our use of a locally measured background.} Comparing the best $C$ statistic from a fit with the temperature, metallicity and redshift free with the best fit when the metallicity is fixed to zero, we find an $F$-test significance of $9.4\E{-3}$, corresponding to a $2.6\sigma$ detection of the emission line.\footnote{To be precise, the model we compare to has its metallicity fixed to zero and its redshift fixed to 1.99 (but a free temperature). The reason for this is that the temperature and redshift are nearly perfectly degenerate for models with zero metallicity. Counting both of these parameters as free when determining the number of degrees of freedom for the $F$ test is therefore questionable. In our case, doing so does not affect the $C$-statistic of the zero-metallicity model (as expected), but would lead to the line-detection significance being estimated as $3.1\sigma$ rather than $2.6\sigma$.} A similar significance is found by brute-force Monte Carlo, i.e.\ generating fake data sets from the best zero-metallicity model and comparing the $\Delta C$ when fitting zero- and free-metallicity models to that obtained from the real data. Note that these calculations account for the fact that the line center is effectively a free parameter to be optimized (because the redshift is free). The right panel of Figure~\ref{fig:spectra} compares the combined EPIC spectrum with the best-fitting model, and shows residuals with respect to a zero-metallicity model with the same temperature.

We also performed a blind search for emission features by fitting a zero-metallicity thermal model plus a Gaussian line profile, where the line energy was allowed to vary from 1.5 to 3.0\,keV (corresponding to redshifts $1.2<z<3.5$ if identified with Fe emission). The preferred solution for this model is an emission line centered at 2.2\,keV, corresponding to the model discussed above. There are two additional local minima in the $C$ statistic, with line energies of 1.8 and 2.6\,keV; however, the line normalization at either of these energies is consistent with zero at 95 per cent confidence, and is nominally negative (i.e., if real they would correspond to absorption features). In contrast, the line normalization when centered at 2.2\,keV is positive at $>99$ per cent confidence, in agreement with the detection significance estimated above.

\begin{figure*}
  \centering
  \includegraphics[scale=\figscale]{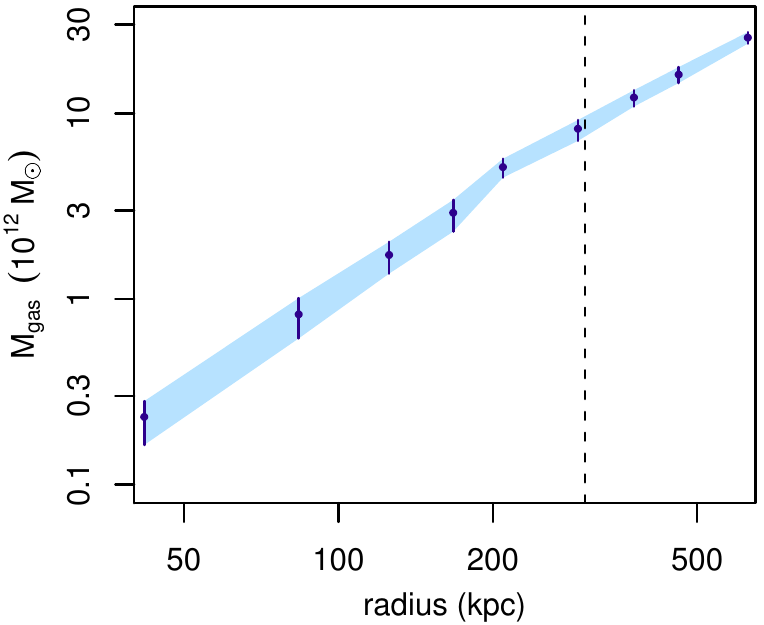}
  \hspace{10mm}
  \includegraphics[scale=\figscale]{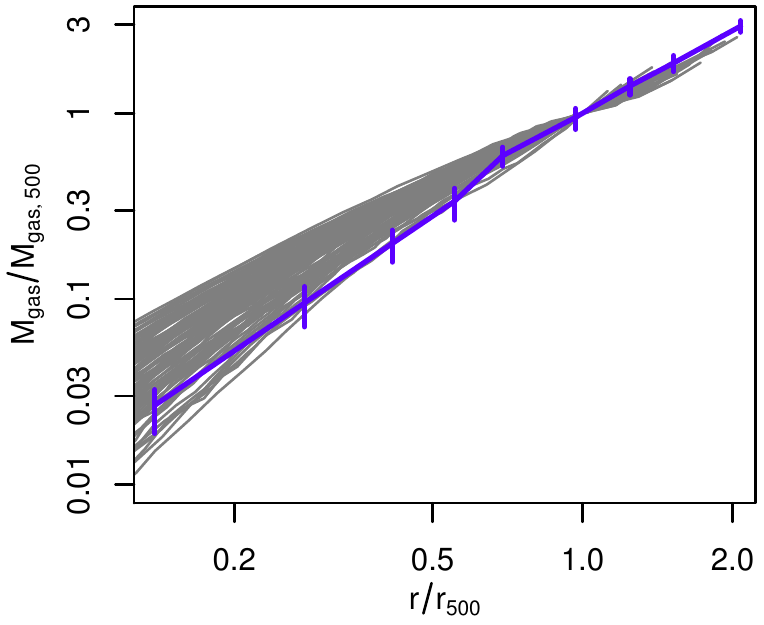}
  \caption{
    Left: gas mass profile of \cl{} as determined from our analysis. Error bars and shading indicate the 68.3 per cent confidence region. The vertical, dashed line shows our estimate of $r_{500}$, determined in Section~\ref{sec:global}.
    Right: the same gas mass profile in scaled units (blue) is compared with an ensemble of scaled profiles for massive, $z<0.5$ clusters from \citet{Mantz1606.03407}.
  }
  \label{fig:Mgas}
\end{figure*}

The posterior distribution for $z$ from fitting the standard thermal model (temperature, metallicity and redshift free) is shown with blue shading in the inset in Figure~\ref{fig:spectra}, and corresponds to $z=1.99^{+0.07}_{-0.06}$, in good agreement with the photometric redshift of $z=1.9^{+0.19}_{-0.21}$ (based on the galaxy redshift histogram shown in the figure; see \citealt{Willis1212.4185}). The X-ray data thus provide a spectroscopic confirmation of the high redshift of this cluster.

Given that the observed spectra at hard energies are completely dominated by the background, we might expect more robust constraints from a fit to the data in a more limited energy band. The difference is small in practice, but we nevertheless adopt as the principal results in this section constraints obtained by fitting only the data at 0.4--3.0\,keV. This fit yields a temperature of $kT=5.0\pm0.7$\,keV, a metallicity in solar units of $Z/\Zsun=0.33^{+0.19}_{-0.17}$, and a redshift of $z=1.99^{+0.07}_{-0.05}$. These values, and the results of the tests discussed above, are listed in Table~\ref{tab:specfits}. Note that, while both temperature and metallicity are degenerate with $z$, this is subdominant to statistical uncertainties in determining the constraints.

\subsection{Gas Mass Profile} \label{sec:deproj}

In order to produce a three-dimensional profile of gas mass in the cluster, we extract spectra in annuli, corresponding to the radial bins shown in Figure~\ref{fig:sbprof}, out to a maximum radius of $75''$ (628\,kpc). As in Section~\ref{sec:spectra}, we use spectra extracted between radii of $2.3'$ and $5'$ to model the background. We model the emissive cluster gas as a series of concentric shells, with each shell corresponding in radius to one of the annuli where source spectra are extracted. The emissivities of each shell are independently free parameters, while the temperature and metallicity of the gas are assumed to be the same in all shells. To compare this model to the data in each annulus, the model spectra from each shell are first mixed according to the geometric projection of the three-dimensional model onto the sky (this part is equivalent to the {\sc projct} model in {\sc xspec}), and then mixed again by the PSF. As in the preceding section, we fit to the data at energies of 0.4--3.0\,keV.

The emissivity constraints in each shell were converted to a gas density profile assuming a canonical value of the mean molecular mass of $\mu=0.61\mproton$, and adopting a fixed redshift of $z=1.99$. We then corrected this profile for projected emission originating at radii (in 3 dimensions) greater than the $75''$ extent of the modelled cluster volume by iteratively fitting a beta model to the tail of the density profile and accounting for the projected emission due to the model continuation; this correction exceeds 0.5 per cent only for our results at radii $\geq 45''$ (377\,kpc) and exceeds the statistical uncertainties only for the outermost point at $75''$ (628\,kpc) radius. The final gas mass profile is shown in the left panel of Figure~\ref{fig:Mgas}.

\begin{figure*}
  \centering
  \includegraphics[scale=\figscale]{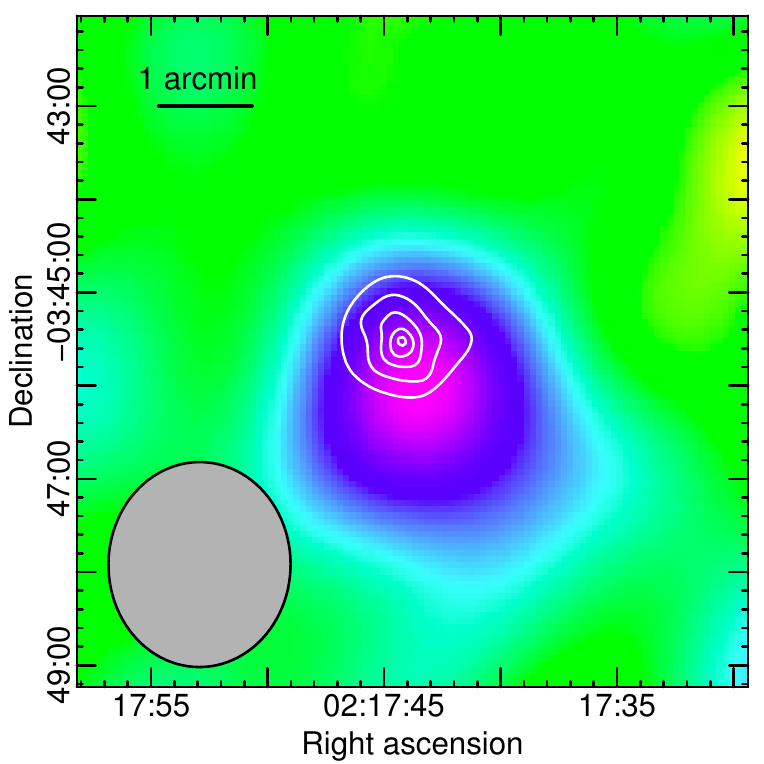}
  \hspace{5mm}
  \includegraphics[scale=\figscale]{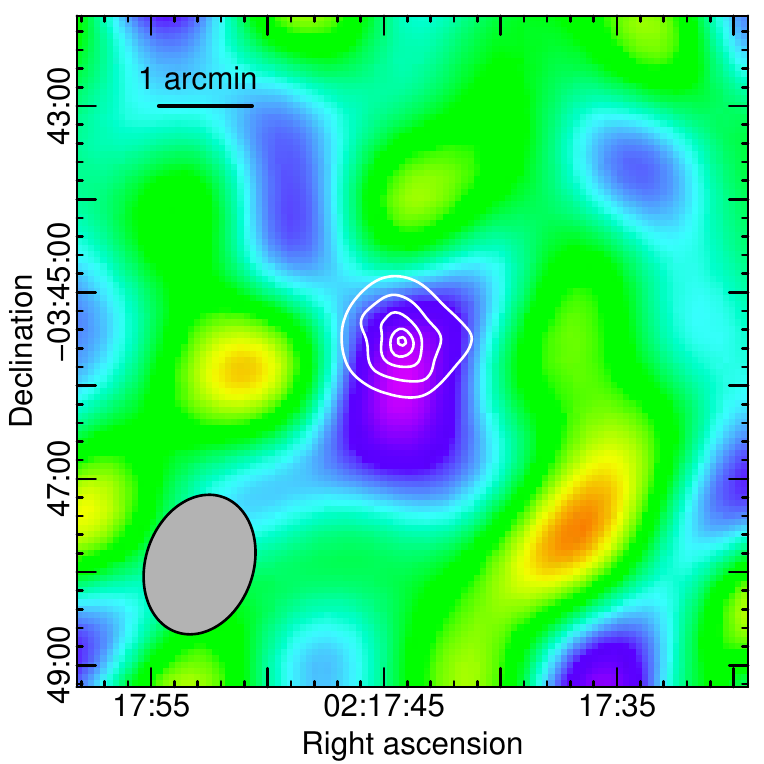}
  \includegraphics[scale=\figscale]{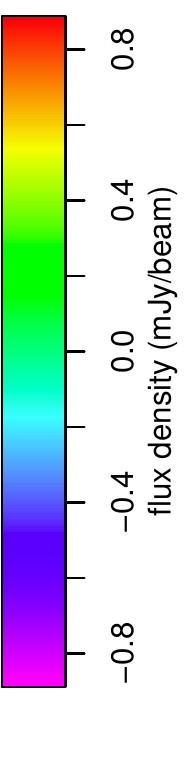}
  \caption{
    Short-baseline ($uv$ radii $<2$k$\lambda$) 30\,GHz maps of \cl{} from the CARMA-8 (left) and CARMA-23 (WB, right) data, after modeling and subtracting point sources and applying the CLEAN image reconstruction algorithm \citep{Hoegbom1974A26AS...15..417}. White contours show the extended X-ray emission, as in Figure~\ref{fig:image}. Gray ellipses in the lower-left corners show the synthesized beam shapes. Both maps use a common color table.
  }
  \label{fig:szimage}
\end{figure*}

\subsection{SZ Signal} \label{sec:sz}

Constraining the SZ effect due to \cl{} requires us to simultaneously model the cluster and any emissive sources in the field. Both types of source can be constrained simultaneously due to the wide range in baseline lengths represented in the CARMA data. Two radio point sources were detected in the original CARMA analysis (XXL Paper~\xxlv{}), at angular separations of $3.6'$ and $5.5'$ from the cluster, and we account for them here. No additional sources are apparent in the full CARMA data set. In particular, there is no evidence for a radio source associated with the BCG of \cl{}, and, if included, its flux is consistent with zero; correspondingly, our results in this section are not affected by whether we model such a source.

Although our quantitative results (see Section~\ref{sec:global}) are fit to the visibilities at all baselines, it is useful to visualize the short-baseline (i.e., cluster-scale) map after fitting and subtracting the point source contribution. Figure~\ref{fig:szimage} shows maps made from the CARMA-8 and CARMA-23 data, using a maximum $uv$ radius of 2\,k$\lambda$. The cluster SZ signal is detected independently in each data set, formally at $6.6\sigma$ and $2.7\sigma$ significance, respectively; the combined detection significance is $7.6\sigma$.

Our procedure for fitting the cluster SZ signal is given in XXL Paper~\xxlv{}. We model the cluster using a generalized NFW (GNFW) form for the three-dimensional, spherically symmetric ICM electron pressure profile, assuming values of the shape parameters given by \citet{Arnaud0910.1234}: $(c_{500},\gamma,\alpha,\beta) = (1.177,0.3081,1.0510,5.4905)$. The remaining cluster parameters are the position of the model center, an overall normalization, and a scale radius, $r_\mathrm{s}=r_{500}/c_{500}$. The Compton $Y$ signal is then obtained, modulo some physical constants, by integrating the electron pressure within a sphere. Since the CARMA data cannot simultaneously constrain the normalization and scale radius of the pressure profile, in Section~\ref{sec:global} we adopt a prior on $r_{500}$ based on the X-ray data in order to measure a value of $Y_{500}$ that is consistent with the ``global'' X-ray measurements.

As noted in XXL Paper~\xxlv{}, the assumed slope of the pressure profile at large radii ($\beta$), which cannot be directly measured from these data, can have a significant influence on the inferred integrated $Y$ parameter. The main results in that work used the pressure profile template obtained from Bolocam data by \citet{Sayers1211.1632}, which had a shallower outer slope than earlier published results \citep{Arnaud0910.1234, Planck1207.4061}. However, a more recent analysis combining Bolocam and \Planck{} data revised this slope to be instead somewhat steeper than those works \citep{Sayers1605.03541}. Here we adopt the \citet{Arnaud0910.1234} template to simplify comparisons with other clusters and because its outer slope is intermediate between the empirical constraints of the \citet{Planck1207.4061} and \citet{Sayers1605.03541}. Differences in the recovered Compton $Y$ among these 3 pressure templates are at the $\sim7$ per cent level, smaller than our statistical uncertainties (Section~\ref{sec:global}). 

A puzzling feature of the SZ signal from \cl{} is that symmetric models like those described above prefer to be centered $\sim35''$ south of the X-ray peak and BCG of the cluster.\footnote{Note that the astrometry of the CARMA data appears good based on the positions of known point sources. This includes the 2\,mJy source present in the cluster field, whose position is consistent with the corresponding 1.4\,GHz FIRST detection within sub-arcsec uncertainties (see XXL Paper~\xxlv{}).} This was noted in XXL Paper~\xxlv{}, and is visually apparent in both the CARMA-8 and CARMA-23 short-baseline maps in Figure~\ref{fig:szimage}. Fitting the combined CARMA data set, we find an offset from the BCG of $35'' \pm 8''$ ($295\pm64$\,kpc at $z=1.99$). Compared to a model whose center is fixed to the BCG position, this has $\Delta \chi^2=-20$, corresponding to $3.8\sigma$ significance. The 68.3 and 95.4 per cent confidence regions for the SZ model center are shown as magenta contours in Figure~\ref{fig:image}; the best-fitting SZ center is J02:17:44.036$-$03:46:06.15. Motivated by the possibility of a merging configuration, we investigated a series of elliptically symmetric and 2-component SZ models, but find that none are statistically preferred by the data. Given the strength of the preference for an offset SZ center, our results for the Compton $Y$ parameter in Section~\ref{sec:global} are based on a fit with the cluster center free, but we note that fixing the model center to the BCG position would reduce the best-fitting $Y_{500}$ value by $\sim13$ per cent, comparable to the statistical uncertainty.

\subsection{Galaxy Profile} \label{sec:galaxies}

\begin{figure}
  \centering
  \includegraphics[scale=\figscale]{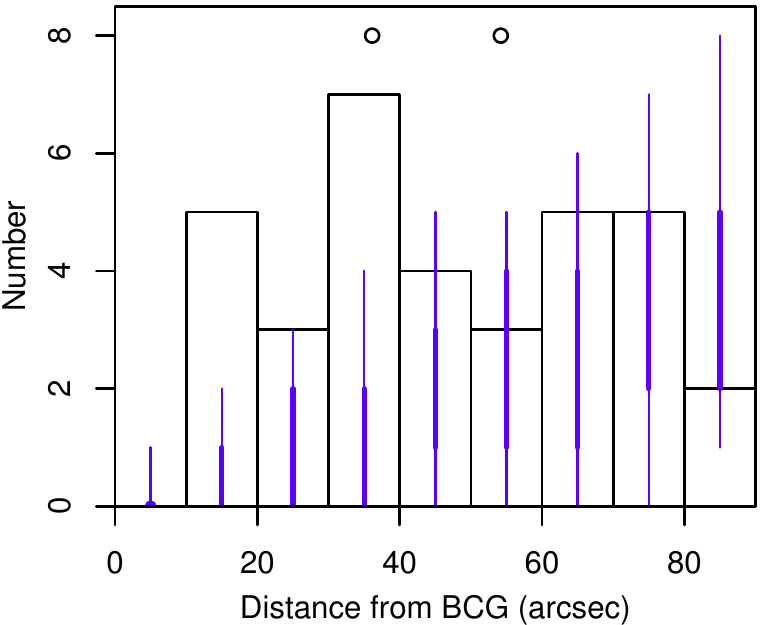}
  \caption{
    Histogram of distance from the BCG for galaxies photometrically placed at redshifts 1.7--2.1 by \citet[][not counting the BCG itself]{Willis1212.4185}. Thick and thin lines overlaid respectively show the $1\sigma$ and $2\sigma$ confidence expectations (reflecting shot noise) based on the background of redshift 1.7--2.1 galaxies in the same observation. Open circles indicate our estimates of $r_{500}$ and $r_{200}$ from Section~\ref{sec:global}.
  }
  \label{fig:gal}
\end{figure}

Although we do not have spectroscopic confirmations of galaxy membership, we can obtain a rough galaxy number profile from the photometric redshift assignments of \citet{Willis1212.4185}. Figure~\ref{fig:gal} shows the number of galaxies with photo-$z$'s in the range 1.7--2.1 as a function of cluster radius, within $90''$ of the BCG position. Thick and thin lines overlaid respectively show the $1\sigma$ and $2\sigma$ confidence expectations (reflecting shot noise) based on the background of redshift 1.7--2.1 galaxies measured from the same IR observations far from the cluster location. There is a clear excess of potential member galaxies within $\sim40''$ of the BCG, corresponding well with the brightest X-ray emission.

\begin{table*}
  \begin{center}
    \caption{
      Global properties of \cl{}.
    }
    \label{tab:global}
    \vspace{-3ex}
    \begin{tabular}{ccccccccccc}
      \hline\hline\vspace{-2ex}\\
      $z$ & $r_{500}$ & $M_{500}$ & $\Mgas$ & $kT$ & $Z$ & $L(0. 1$--$2.4\keV)$ & $L(0.5$--$2.0\keV)$ & $Y_{500}$ & $Y_{500}\,\dA^2$\vspace{0.3ex}\\
      & (kpc) & $(10^{13}\Msun)$ & $(10^{12}\Msun)$ & (keV) & $(\Zsun)$ & $(10^{44}\erg\second^{-1})$ & $(10^{44}\erg\second^{-1})$ & $(10^{-12})$ & $(10^{-5}\,\Mpc^2)$\vspace{0.2ex}\\
      \hline\vspace{-2ex}\\
      $1.99^{+0.07}_{-0.05}$ & $295\pm23$ &  $6.3\pm1.5$ & $7.9\pm1.9$ & $5.0\pm0.7$ & $0.33^{+0.19}_{-0.17}$ & $3.5\pm0.5$ & $2.2\pm0.3$ & $3.6\pm0.4$ & $1.07\pm0.13$\vspace{0.5ex}\\
      \hline
   \end{tabular}
   \tablefoot{
     Where appropriate, measurements are referenced to the characteristic radius $r_{500}$. A redshift of $z=1.99$ is assumed in the derivations of mass, gas mass, luminosity and Compton $Y$. The impact of the redshift uncertainty on other quantities (e.g.\ through the angular diameter distance) is subdominant to statistical uncertainties.
   }
  \end{center}
\end{table*} 

\subsection{Global Properties and Mass Estimate} \label{sec:global}

For sufficiently massive clusters, the gas mass fraction at intermediate-to-large radii is expected to be approximately constant, as simulations (including those that implement hydrodynamic heating and cooling processes) have verified (e.g.\ \citealt{Eke9708070}; \citealt*{Nagai0609247}; \citealt{Battaglia1209.4082, Planelles1209.5058, Barnes1607.04569}). Our gas mass profile can therefore be used to provide an estimate of the mass of \cl{} that is arguably more secure than those based on extrapolating other scaling relations (all of which are necessarily calibrated at significantly lower redshifts) out to $z\sim2$. We adopt the fiducial value of $\fgas(r_{500})=0.125$, based on results from massive, X-ray selected clusters at $z<0.5$ (and consistent with dynamically relaxed clusters at redshifts $z<1.06$; \citealt{Mantz1509.01322, Mantz1606.03407}).\footnote{Applying a typical mass accretion history (e.g.\ \citealt{McBride0902.3659}) to our estimate of $M_{500}\sim6\E{13}\Msun$ for \cl{} suggests that it will grow into an $M_{500}\sim2$--$6\E{14}\Msun$ cluster by $z=0$. This range overlaps well with the sample used to calibrate our reference value of $\fgas$ at low redshift ($M_{500}\gtsim4\E{14}\Msun$; \citealt{Mantz1606.03407}), indicating that we can consistently make use of a gas fraction appropriate for cluster-scale halos. We note that extrapolating scaling relations for luminosity, temperature or $\Yx$ would typically lead to a larger total mass estimate (see Figures~\ref{fig:scaling} and \ref{fig:scaling-spt}, and estimates in XXL Paper~\xxlv{}).} Note that this $\fgas$ value follows from total masses measured from weak lensing shear \citep{Applegate1208.0605, Applegate1509.02162} and gas masses from \Chandra{} (which are generally found to be in good agreement with XMM gas masses; \citealt{TsujimotoAA...525A..25, Rozo1204.6301, Schellenberger1404.7130}).

We arrive at an estimate of $M_{500}$, and the corresponding radius $r_{500}$, by solving the implicit equation
\begin{equation} \label{eq:rDelta}
  M(r_{500}) = \frac{\Mgas(r_{500})}{\fgas(r_{500})} = \frac{4\pi}{3} \, 500\rhocr(z) r_{500}^3.
\end{equation}
Propagating the uncertainties in the $\Mgas(r)$ profile forward, we find $r_{500}=295\pm23$\,kpc ($35''\pm3''$) and $M_{500}=(6.3\pm1.5)\E{13}\Msun$; correspondingly, $\Mgas{}_{,500}=(7.9\pm1.9)\E{12}\Msun$.\footnote{In this procedure, we have assumed a fixed redshift of $z=1.99$, but we note that, within the redshift constraint provided by the data, the dependence of critical density and angular diameter distance on $z$ are subdominant to statistical uncertainties. The same applies to the constraints on luminosity and Compton $Y$, below.} The right panel of Figure~\ref{fig:Mgas} compares our gas mass profile, scaled in units of $r_{500}$ and $\Mgas{}_{,500}$, with similarly scaled profiles from \citet{Mantz1606.03407}; the similarity in shape between the profile of \cl{} and those of the massive, $z<0.5$ clusters is encouraging. Assuming a factor of $\approx 1.5$ relating $r_{500}$ with $r_{200}$ \citep{Navarro9611107}, we have $r_{200}=443\pm35$\,kpc ($53''\pm3''$).

The emissivity profile fit from Section~\ref{sec:deproj} can be used to determine the unabsorbed, rest-frame cluster luminosity in a given energy band projected within $r_{500}$, accounting for the statistical uncertainties in the gas temperature. In two commonly used reference bands, we find $L(0. 1$--$2.4\keV)=(3.5\pm0.5)\E{44}\erg\second^{-1}$ and $L(0.5$--$2.0\keV)=(2.2\pm0.3)\E{44}\erg\second^{-1}$. Given that the constraint on $r_{500}$ above is in good agreement with the outer radius used for the spectral analysis of Section~\ref{sec:spectra}, we take the temperature and metallicity results from that section as appropriate for the region $r<r_{500}$. Incorporating the X-ray $r_{500}$ constraint as a prior when fitting the CARMA SZ data, we find the spherically integrated Compton parameter within $r_{500}$ to be $Y_{500}=(3.6\pm0.4)\E{-12}$, or $Y_{500}\,\dA^2(z)=(1.07\pm0.13)\E{-5}\,\Mpc^2$ for $z=1.99$ (where $\dA$ is the angular diameter distance). All of these characteristic properties of the cluster are collected in Table~\ref{tab:global}.

\section{Discussion} \label{sec:discussion}

\subsection{Comparison with Lower-Redshift Samples} \label{sec:scaling}

\begin{figure*}
  \centering
  \includegraphics[scale=\figscale]{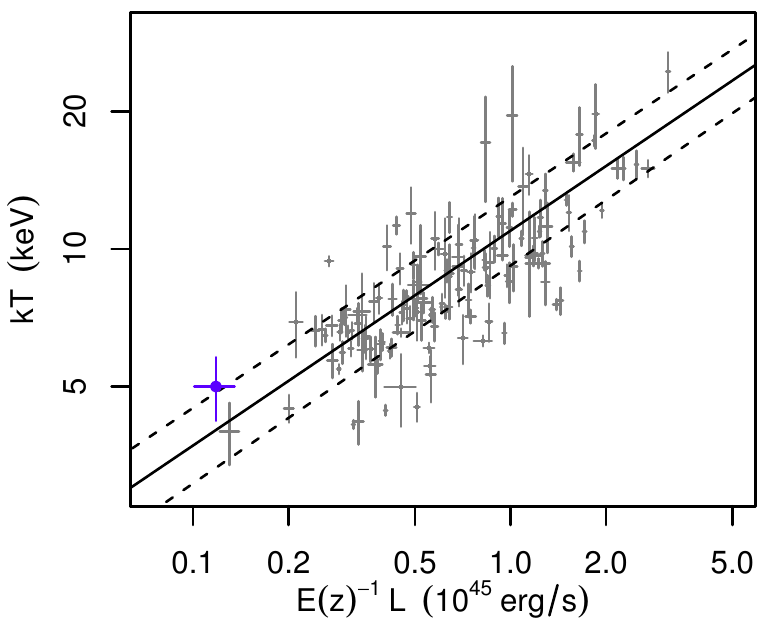}
  \hspace{10mm}
  \includegraphics[scale=\figscale]{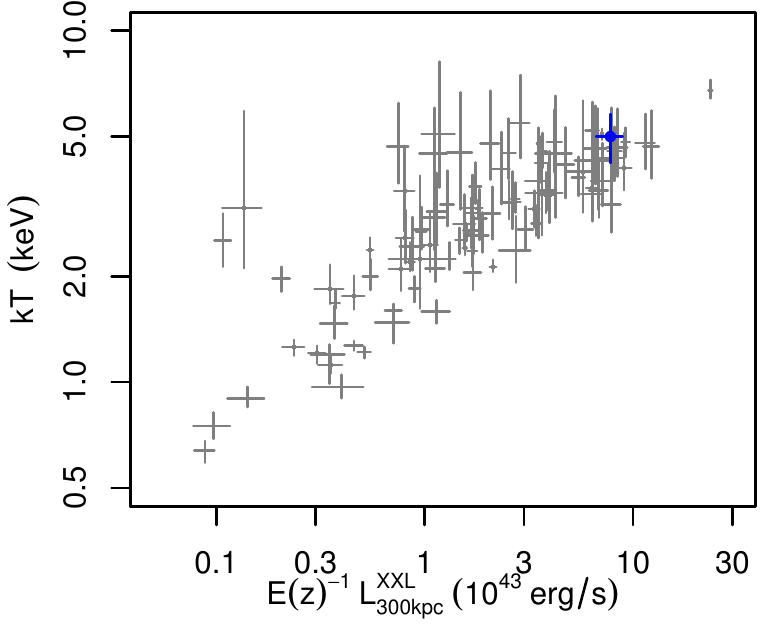}
  \bigskip\\
  \includegraphics[scale=\figscale]{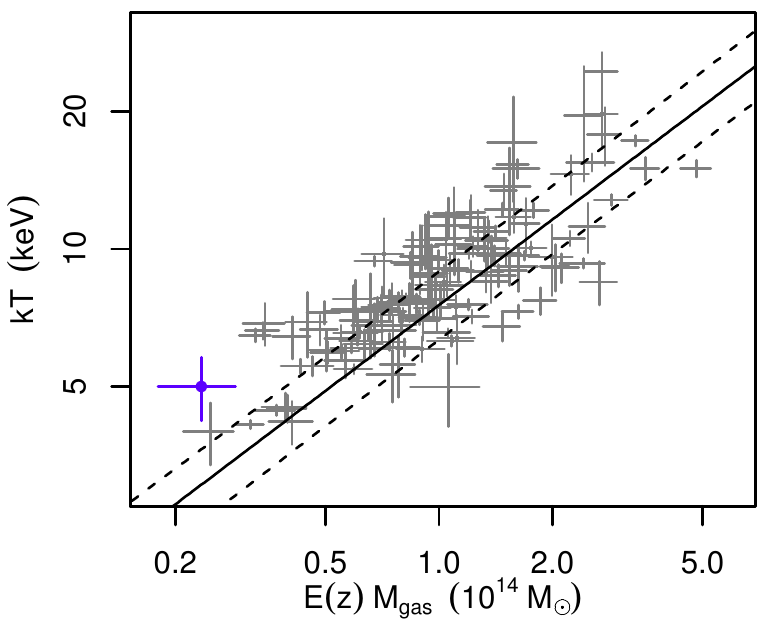}
  \hspace{10mm}
  \includegraphics[scale=\figscale]{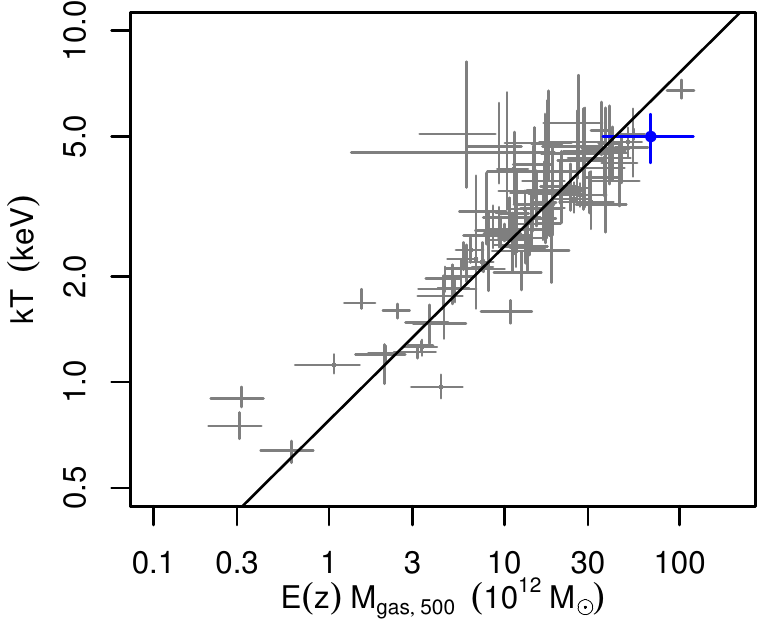}
  \bigskip\\
  \includegraphics[scale=\figscale]{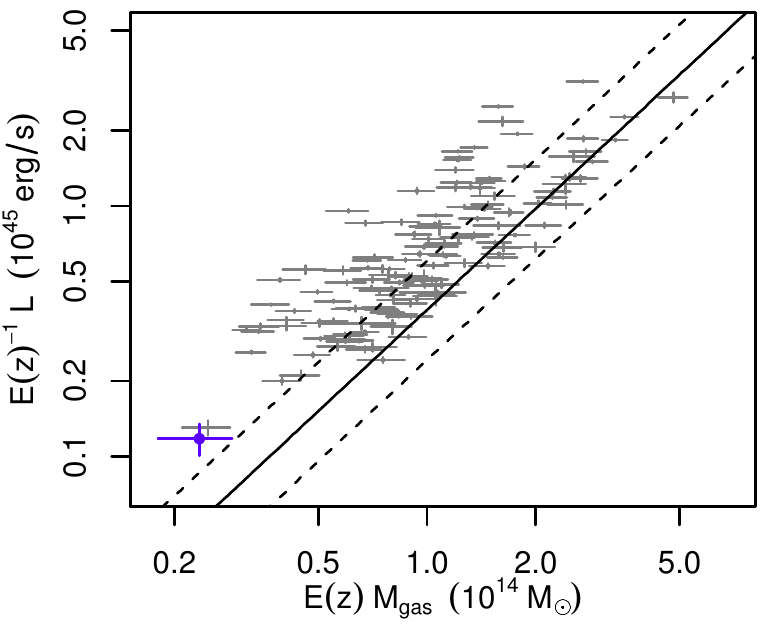}
  \hspace{10mm}
  \includegraphics[scale=\figscale]{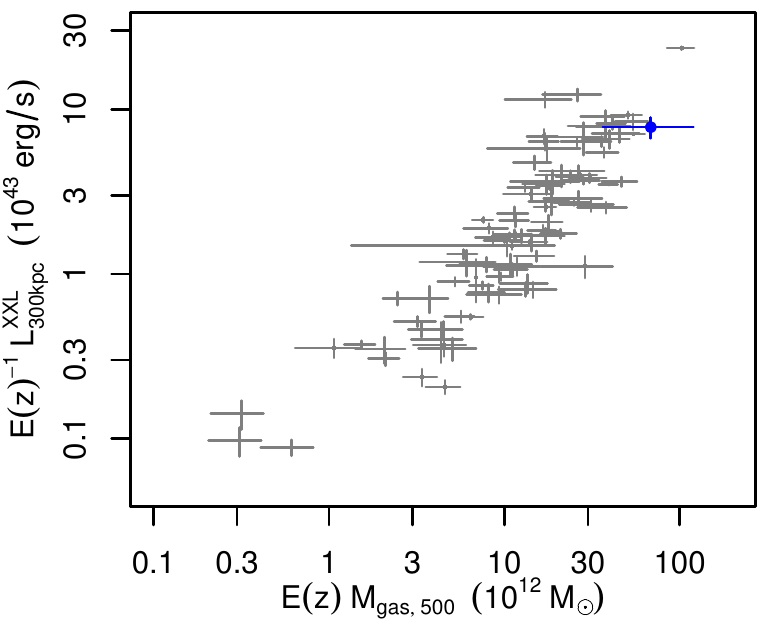}
  \caption{
    Comparison of the global properties of \cl{} (blue point) with lower-$z$, X-ray selected cluster samples. Factors of $E(z)=H(z)/H_0$ encode self-similar evolution of the scaling relations.
    Left column: gray points show measurements from \citet[][$z<0.5$]{Mantz1606.03407}, and solid/dashed lines the corresponding scaling relations (accounting for X-ray flux-selection bias) and their uncertainty (including intrinsic scatter). These measurements were obtained from \Chandra{} data, but otherwise the procedure for determining each observable (and $r_{500}$) is essentially the identical to the one used in this work. Luminosities correspond to the rest-frame 0.1--2.4\,keV band. Note that temperatures from \citet{Mantz1606.03407} are center-excised (excluding radii $<0.15\,r_{500}$), while the measurement for \cl{} is not.
    Right column: gray points show measurements of the 100 brightest XXL clusters ($z<1.05$; XXL Papers~\xxlii{}, \xxliii{} and \xxlxiii{}). Luminosities correspond to the rest-frame 0.5--2.0\,keV band, and luminosities and temperatures are measured in an aperture of radius 300\,kpc. Gas masses are measured within $r_{500}$, as estimated from the mass--temperature relation of XXL Paper~\xxliv{}. In the center-right panel, solid line shows the $\Mgas$--$T$ relation of XXL Paper~\xxlxiii{}.
  }
  \label{fig:scaling}
\end{figure*}
 
In this section, we compare the global properties of \cl{} with those of X-ray and SZ selected cluster samples at lower redshifts. We first compare to the sample of \citet{Mantz1606.03407}, which comprises the most massive, X-ray selected clusters at $0.0<z<0.5$, and thus provides a long lever arm in redshift. Another advantage of using this sample is that the procedure used to determine the global X-ray properties is nearly identical to the one employed here. In particular, the method for estimating $r_{500}$ from the gas mass profile is identical. The principal difference between the analyses is the use of data from XMM rather than \Chandra{} in the present work, and the consequent need to account for the effect of the PSF. We note that the temperature we obtain for \cl{} is likely to be comparable to temperatures measured by \Chandra{} despite the well publicized disagreements between the two telescopes at high temperatures; this is because the cluster emission in this case is redshifted to observer-frame energies $\ltsim 2.5$\,keV, where the instrumental responses of the two observatories are in good agreement \citep{TsujimotoAA...525A..25, Schellenberger1404.7130}. Another issue is the use of center-excised temperature by \citet[][specifically, excising radii $<0.15\,r_{500}$]{Mantz1606.03407}, which we cannot easily replicate due to the small angular extent of \cl{} compared with the XMM PSF. However, as we do not expect well developed cool cores to exist in clusters at $z\sim2$, this is also likely to have a small impact on the comparison.

\begin{figure*}
  \centering
  \includegraphics[scale=\figscale]{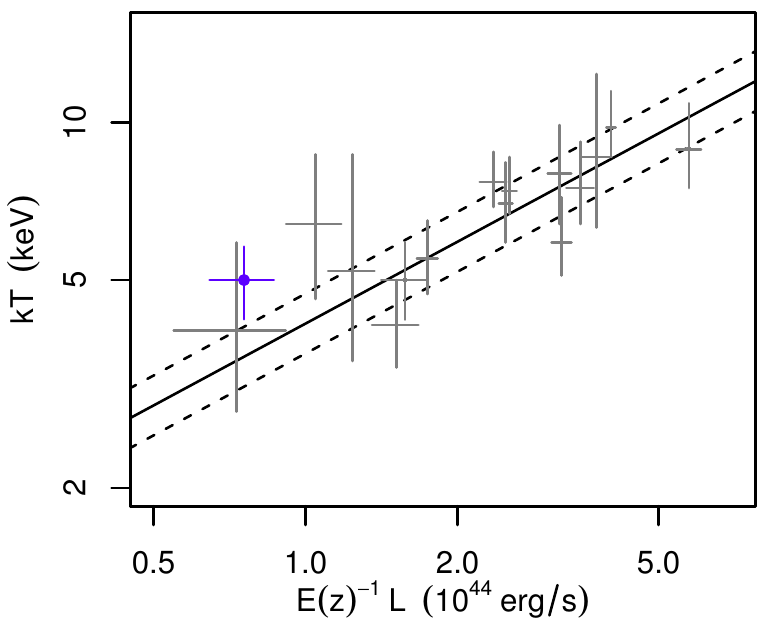}
  \hspace{10mm}
  \includegraphics[scale=\figscale]{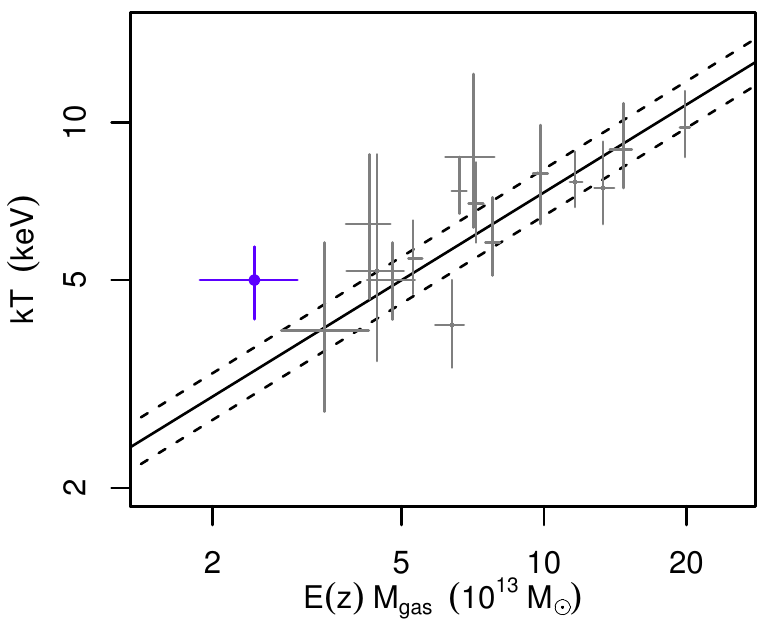}
  \bigskip\\
  \includegraphics[scale=\figscale]{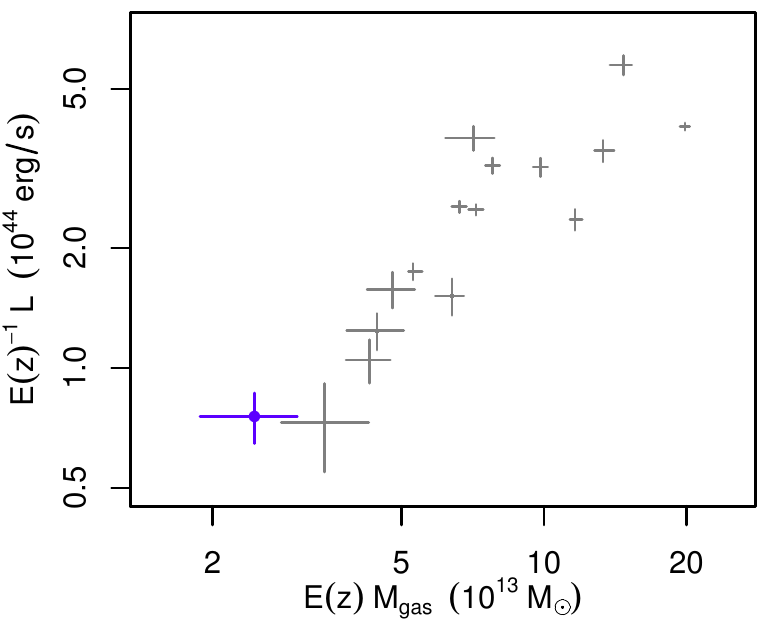}
  \hspace{10mm}
  \includegraphics[scale=\figscale]{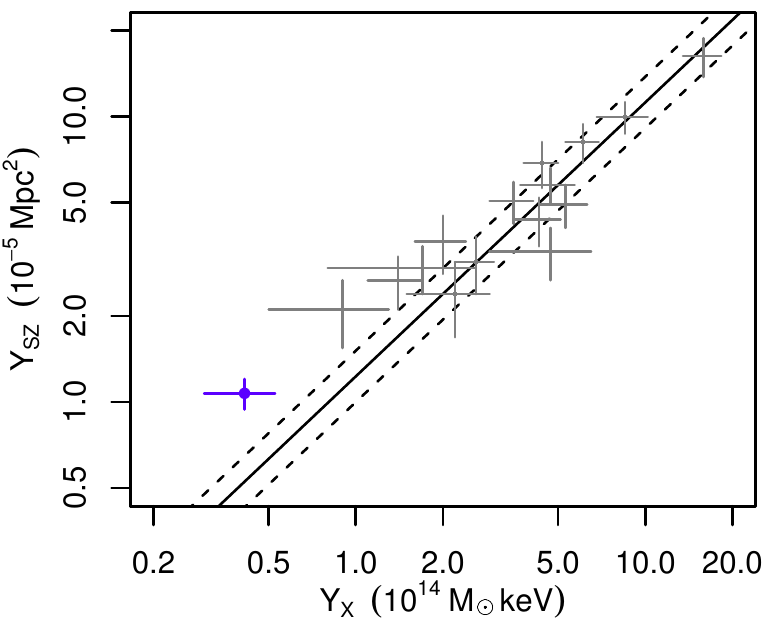}
  \caption{
    Comparison of the global properties of \cl{} (blue point) with lower-$z$, SZ-selected clusters from SPT at redshifts $0.32<z<1.07$ \citep[gray points;][]{Andersson1006.3068}. Luminosities are in the 0.5--2.0\,keV rest-frame band. Note that the SPT cluster temperatures are center-excised (excluding radii $<0.15\,r_{500}$), while the measurement for \cl{} is not. Solid and dashed lines show the scaling relation constraints published by \citet{Andersson1006.3068}. Factors of $E(z)=H(z)/H_0$ encode self-similar evolution of the scaling relations.
  }
  \label{fig:scaling-spt}
\end{figure*}
 
The left column of Figure~\ref{fig:scaling} compares our measurements of gas mass, temperature, and 0.1--2.4\,keV intrinsic luminosity for \cl{} (blue points) with the \citet{Mantz1606.03407} $z<0.5$ sample (gray points). Also shown are the scaling relations derived in that work, with the width indicated by the dashed lines primarily reflecting the intrinsic scatter in those relations. Considering both the statistical uncertainties and the intrinsic scatter, the agreement is good, despite the significant difference in redshift between the two data sets.

We next compare with the sample of 100 brightest (in flux) XXL-detected clusters (XXL Paper~\xxlii{}), of which \cl{} is not a member. The 100-brightest XXL sample spans redshifts of 0.04--1.05 and 0.5--2.0\,keV luminosities of $\sim9\E{41}$--$3\E{44}\erg\second^{-1}$. \citet[][hereafter XXL Paper~\xxliii]{Giles1512.03833} report temperatures and 0.5--2.0\,keV luminosities measured in an aperture of 300\,kpc radius for each cluster, similar to the our estimate of $r_{500}$ for \cl{}; this comparison is shown in the top-right panel of Figure~\ref{fig:scaling}. The middle- and bottom-right panels compare the same luminosity and temperature measurements with gas masses from \citet[][hereafter XXL Paper \xxlxiii{}]{Eckert1512.03814}. For consistency with these authors, the gas mass for \cl{} has been recomputed in these plots, using a value of $r_{500}=(406\pm81)$\,kpc estimated from the mass--temperature relation of \citet[][hereafter XXL Paper~\xxliv{}]{Lieu1512.03857}. Again, the agreement with a significantly lower-redshift data set is good.

The final comparison, shown in Figure~\ref{fig:scaling-spt}, is to the SZ-selected sample of South Pole Telescope (SPT) clusters published by \citet{Andersson1006.3068}. This sample spans $0.32<z<1.07$, and has measured X-ray properties from \Chandra{} or XMM. In addition to X-ray luminosity (shown in the 0.5--2.0\,keV band), temperature and gas mass, for this sample we can form scaling relations involving the Compton $Y$ parameter; here we show the $Y_{500}$--$\Yx$ relation, where $\Yx=\Mgas{}_{,500}\,kT$. The agreement of \cl{} with the scaling relations derived from this cluster sample, now selected in an entirely different way to the previous cases, is again broadly good when considering the statistical and intrinsic scatters.

Note that self-similar evolution factors \citep{Kaiser1986MNRAS.222..323K} have been applied to luminosity and gas mass in all of these plots. While we cannot draw strong conclusions based on a single cluster, the results in this section suggest that departures from self-similar evolution may be relatively mild for massive clusters, even out to $z=2$.

\subsection{Metallicity}

While the statistical uncertainties on the metallicity of \cl{} from Section~\ref{sec:spectra}, $Z/\Zsun=0.33^{+0.19}_{-0.17}$, are not particularly constraining, the nominal metallicity is tantalizingly similar to values measured in massive clusters at much lower redshifts (excepting the cores of cool core clusters, where star formation is relatively efficient; e.g.\ \citealt{Leccardi0806.1445, Maughan0703156, Baldi1111.4337, Ettori1504.02107, McDonald1603.03035}). This is particularly interesting in light of measurements of extremely uniform metallicities extending to very large radii in the Perseus and Virgo Clusters \citep{Werner1310.7948, Simionescu1506.06164}. Such a distribution is most easily explained by an early enrichment scenario, in which the intergalactic medium is enriched to a metallicity of $Z/\Zsun\approx0.3$ prior to the formation of massive clusters, at $z>2$--3. This picture is reinforced by recent \Chandra{} measurements showing the metallicity at intermediate cluster radii (core-excised) to be consistent with a constant out to $z\approx1.5$ (\citealt{McDonald1603.03035}). The measured metallicity presented here, if confirmed by more precise X-ray measurements for \cl{} and similarly massive, high-redshift clusters, would provide a definitive confirmation of the enrichment of the ICM at early times.

\subsection{Dynamical State}

At first glance, the dynamical state of \cl{} presents something of a paradox. The north-south offset between the X-ray peak/centroid and the SZ center (Section~\ref{sec:sz}) is large compared with to the extent of the cluster, $(0.99\pm0.23) r_{500}$ (including measurement uncertainties in both the offset and $r_{500}$). At face value, this suggests significant thermodynamic asymmetry. In simulations, \citet{Zhang1406.4019} find that X-ray/SZ offsets of several hundred kpc are possible in merging configurations similar to that of the Bullet Cluster \citep{Markevitch0110468}, where one subcluster contains significantly X-ray brighter gas than the other. In such a case, we would still expect to see extended X-ray emission associated with a lower surface brightness subcluster, coincident with the SZ center. The only diffuse X-ray emission detectable in the current data appears well centered on the X-ray peak (note that this emission is more extended than the PSF; Figure~\ref{fig:sbprof}). The visible emission is, however, slightly elongated in the direction of the SZ detection (Figure~\ref{fig:image}).

We note that the interpretation of the measured offset in the context of the work of \citet{Zhang1406.4019} is not entirely straightforward, since those authors define the SZ center as the peak of a relatively high-resolution ($2''$--$15''$) map. The correspondance between this peak and the center of a symmetric model fit to our interferometric data is non-trivial in the case of an asymmetric cluster. In this context, a useful, empirical comparison can be made with the X-ray/SZ offsets measured by \citet{Andersson1006.3068} for 15 SPT clusters. The SPT data is natively of similar resolution to our CARMA data, and the clusters are detected with comparable signal-to-noise to \cl{}; the SZ centering method used (maximizing the significance of a beta-model cluster profile) is also broadly similar to ours. The SPT X-ray/SZ offsets in angular separation, the most relevant comparison in terms of resolution and centroiding accuracy, are similar in magnitude to the value we measure; our offset of $35''$ lies at the 87th percentile of the \citet{Andersson1006.3068} sample (i.e.\ two of their clusters are more extreme). In terms of metric distance, \cl{} corresponds to the 93rd percentile (1 SPT cluster more extreme).

In summary, the measured X-ray/SZ offset for \cl{} is relatively large, though within the range observed for other clusters, and could be explained by a merging configuration. At the same time, the diffuse X-ray emission visible in the current data, taken in isolation, appears regular, and is well aligned with the putative BCG. Higher spatial resolution X-ray and/or SZ data, revealing any small scale structure in the ICM, would provide the most straightforward means to address this question.

\section{Conclusion} \label{sec:conclusion}

We present results from a 100\,ks XMM observation of galaxy cluster \cl{}, the first massive cluster discovered through its X-ray emission at $z\approx2$. The data allow us to, for the first time, measure global thermodynamic properties such as temperature, metallicity and gas mass for such a system. The rest-frame 6.7\,keV Fe emission complex is detected in the EPIC data, providing an emission-weighted metallicity constraint of $Z/\Zsun=0.33^{+0.19}_{-0.17}$. The Fe emission line detection furthermore directly and spectroscopically confirms the high redshift of this cluster, $z=1.99^{+0.07}_{-0.06}$, in agreement with the earlier photometric estimate of $1.91^{+0.19}_{-0.21}$. The temperature constraint of $kT=5.0\pm0.7$\,keV confirms that \cl{} is indeed a massive cluster, containing a hot ICM. Accounting for the PSF, we generate a gas mass profile for \cl{}, and use this to estimate a characteristic radius of $r_{500}=(295\pm23)$\,kpc (corresponding to a mass of $M_{500}=(6.3\pm1.5)\E{13}\Msun$), assuming a constant gas mass fraction at $r_{500}$ of 0.125. We additionally provide measurements of the X-ray luminosity and the Compton $Y$ parameter, the latter from CARMA SZ data.

These global properties of \cl{} are in reasonably good agreement with measurements for large samples of clusters in various (lower) redshift ranges (spanning overall $0\ltsim z\ltsim1$) and with scaling relations fitted to those samples, when assuming self-similar evolution. While broad conclusions should not be drawn from a single high-redshift cluster, this good agreement suggests that, for sufficiently massive clusters, the evolution of the ICM is remarkably simple, even out to the highest redshifts. Similarly, the global metallicity measured from our XMM data is in excellent agreement, albeit with large statistical uncertainties, with lower-redshift clusters, supporting a picture in which the ICM throughout most of the cluster volume is enriched at yet higher redshifts. Observations of additional clusters at comparable redshifts will be required to determine how representative, or not, \cl{} is. While \cl{} is a rare object, the prospects for doing cluster statistics at very high redshift are good, with SZ and IR surveys of thousands of square degrees now routinely discovering clusters out to redshifts $z\sim1.7$; as these efforts scale up to larger areas, useful numbers of clusters at even higher redshifts will be uncovered.

In spite of its apparently simple scaling properties, \cl{} presents a puzzle in the spatial offset between the centers of the X-ray and SZ signals. This speaks to the challenge of understanding the internal structure of very distant clusters, given that current X-ray observatories do not simultaneously provide high spatial resolution and high throughput. The SZ effect may provide a useful alternative for assessing the morphology of high-$z$ clusters, either through interferometry (e.g.\ ALMA) or large, single-dish telescopes coupled to sensitive detector arrays. Looking farther ahead, the epoch of cluster formation is one for which large-area, high-resolution X-ray facilities such as Athena and Lynx are ideally suited.

\begin{acknowledgements}
  XXL is an international project based around an XMM Very Large Programme surveying two 25\,deg$^2$ extragalactic fields at a depth of $\sim5\E{15}\erg\cm^{-2}\second^{-1}$ in the 0.5--2\,keV band for point-like sources. The XXL website is \url{http://irfu.cea.fr/xxl}. Multi-band information and spectroscopic follow-up of the X-ray sources are obtained through a number of survey programmes, summarised at \url{http://xxlmultiwave.pbworks.com/}.

\newline\indent We acknowledge support from the National Aeronautics and Space Administration under Grant No.\ NNX16AH27G, issued through the XMM-{\it Newton} Guest Observer Facility, Chandra Award Number GO6-17116A, issued by the Chandra X-ray Observatory Center, and Grant no.\ NNX15AE12G, issued through the ROSES 2014 Astrophysics Data Analysis Program; and from the U.S. Department of Energy under contract number DE-AC02-76SF00515.
  BJM acknowledges support from UK Science and Technology Facilities Council (STFC) grant ST/M000907/1. CHAL acknowledges support from an STFC postgraduate studentship.

\newline\indent Support for CARMA construction was derived from the states of California, Illinois, and Maryland, the James S.\ McDonnell Foundation, the Gordon and Betty Moore Foundation, the Kenneth T.\ and Eileen L.\ Norris Foundation, the University of Chicago, the Associates of the California Institute of Technology, and the National Science Foundation (NSF). CARMA development and operations were supported by the NSF under a cooperative agreement, and by the CARMA partner universities.
\end{acknowledgements}

\def \gca {Geochim.\ Cosmochim.\ Acta}
\def \aap {A\&A} 
\def \apj {ApJ}
\def \apjs {ApJS}
\def \apjl {ApJ} 
\def \mnras {MNRAS}

\end{document}